\documentclass[preprint,one-column,nofootinbib,english]{revtex4-1}
\usepackage{amsthm,amsmath,amsfonts,dsfont,upgreek} %bbm,bbold,,txfonts,mathbbol
\usepackage{amssymb,epsfig,setspace}
\usepackage{graphicx}
\usepackage{babel}
\usepackage{verbatim}
\usepackage{mathrsfs}
\usepackage{enumitem}
\usepackage{bm}% bold math
\usepackage{color}
\usepackage{xcolor}
\def\op#1{{\Hat{\mathrm{#1}}}}

\begin{document}

\hyphenation{ano-ther ge-ne-ra-te dif-fe-rent know-le-d-ge po-ly-no-mi-al}
\hyphenation{me-di-um or-tho-go-nal as-su-ming pri-mi-ti-ve pe-ri-o-di-ci-ty}
\hyphenation{mul-ti-p-le-sca-t-te-ri-ng i-te-ra-ti-ng e-q-ua-ti-on}
\hyphenation{wa-ves di-men-si-o-nal ge-ne-ral the-o-ry sca-t-te-ri-ng}
\hyphenation{di-f-fe-r-ent tra-je-c-to-ries e-le-c-tro-ma-g-ne-tic pho-to-nic}
\hyphenation{Ray-le-i-gh di-n-ger Kra-jew-ska Wal-czak Ham-bur-ger Ad-di-ti-o-nal-ly}
\hyphenation{Kon-ver-genz-the-o-rie ori-gi-nal in-vi-si-b-le cha-rac-te-ri-zed}
\hyphenation{sa-ti-s-fy}
\hyphenation{Ne-ver-the-less sa-tu-ra-te E-ner-gy}

\title{A unified treatment of polynomial solutions and constraint polynomials
of the Rabi models}

\author{Alexander Moroz}

\affiliation{Wave-scattering.com}
%%\end{array}d{wavescattering@yahoo.com}
%%\end{array}d[url]{www.wavescattering.com} 
 
%%\date{\today}
\begin{abstract}
General concept of a gradation slicing is used to analyze polynomial 
solutions of ordinary differential equations (ODE) with polynomial
coefficients, ${\cal L}\psi=0$, where ${\cal L}=\sum_l p_l(z) d_z^l$, 
$p_l(z)$ are polynomials, $z$ is a one-dimensional coordinate, and $d_z=d/dz$. 
It is not required that ODE is either (i) Fuchsian or (ii) leads to a usual 
Sturm-Liouville eigenvalue problem. General necessary and sufficient conditions 
for the existence of a polynomial solution are formulated involving constraint relations. 
The necessary condition for a polynomial solution of $n$th degree to exist 
forces energy to a $n$th baseline. Once the constraint relations on the $n$th baseline 
can be solved, a polynomial solution is in principle possible even in the absence of any 
underlying algebraic structure. The usefulness of theory is demonstrated on the examples of 
various  Rabi models. For those models, a baseline is known as a Juddian baseline (e.g. in the case of the Rabi model 
the curve described by the $n$th energy level of a displaced harmonic oscillator with varying coupling $g$). 
The corresponding constraint relations are shown to (i) reproduce known constraint
polynomials for the usual and driven Rabi models and (ii) generate hitherto
unknown constraint polynomials for the two-mode, two-photon, and generalized 
Rabi models, implying that the eigenvalues of corresponding polynomial eigenfunctions 
can be determined algebraically. Interestingly, the ODE of the above Rabi models are shown to 
be characterized, at least for some parameter range, by the same unique set of grading parameters.
\end{abstract}

%% \begin{keyword}
%% Rabi model \sep Schweber's quantization criterion \sep energy levels 
%% \sep solvability \sep integrability
%% \end{keyword}
%% \end{frontmatter}

%\pacs{03.65.Ge, 02.30.Ik, 42.50.Pq}
%\vspace*{0.9cm}
%
%\newpage
%\setcounter{page}{1}

\maketitle

\section{Introduction}
\label{sc:intr}
%%%%%%%%%%%%%%%%%%%%%%%%%%%%%%%%%%%%%
The Rabi model \cite{Rb} describes the simplest interaction between a 
cavity mode with a bare frequency $\omega$ and a two-level system 
with the levels separated by a frequency difference $2\Delta=\omega_0$,
where $\hbar$ is the Planck constant and $\omega_0$ is a bare resonance frequency. 
The model is characterized by the Hamiltonian \cite{Rb,Schw} 
\begin{equation}
\hat{H}_R =
\hbar \omega \mathds{1} \hat{a}^\dagger \hat{a} 
 + \hbar g\sigma_1 (\hat{a}^\dagger + \hat{a}) + \hbar \Delta \sigma_3,
\label{rabih}
\end{equation}
where $\mathds{1}$ is the unit matrix, $\hat{a}$ and $\hat{a}^\dagger$ are 
the conventional boson annihilation and creation operators 
of a boson mode with frequency $\omega$, which satisfy commutation relation 
$[\hat{a},\hat{a}^{\dagger}] = 1$, and $g$ is a coupling constant. 
Here and elsewhere the standard representation of the Pauli matrices 
$\sigma_l$, $l=1,2,3$, with $\sigma_3$ diagonal is assumed.
The Rabi model applies to a great variety of physical systems, 
including cavity and circuit quantum electrodynamics, quantum
dots, polaronic physics and trapped ions \cite{KGK,BGA,NDH,TAP,TPG,AMdn}. 
The Rabi model is {\em not} exactly solvable. Yet the model has been known for a long
time to possess polynomial solutions, the so called Juddian isolated exact solutions \cite{Jd,Ks},
at energy levels corresponding to those of a {\em displaced harmonic 
oscillator}, which is the $\Delta=0$ limit of the Rabi model \cite{Schw}.
The latter has been known as the {\em baseline} condition for the Rabi model \cite{Schw,Jd,Ks}
[e.g. Eqs. (\ref{c+cn}), (\ref{Rabi-solution-E}), (\ref{2-photon-energy}), 
(\ref{2-mode-energy}), (\ref{gRnc}) below].

In what follows we shall consider the (driven) Rabi model \cite{Zh2,Zh6,Li15,Wa,KRW}, together with 
its nonlinear two-photon \cite{AMdn,Zh2,Zh6,NLL,EB1}
and nonlinear two-mode \cite{AMdn,Zh2,Zh6} versions, and the generalized Rabi model of Refs. \cite{TAP,TPG,AMdn}.
A typical $2$nd order linear ordinary differential equation (ODE)
for the Rabi models turns out to be of the form ${\cal L}\psi=0$, where ${\cal L}=\sum_l p_l(z) d_z^l$, 
$p_l(z)$ are polynomials, $z$ is a one-dimensional coordinate, and $d_z=d/dz$.
What sets those equations apart from other common equations is that ${\cal L}$ comprises energy 
$E$ dependent terms $\sim Ez d_z$, $Ez$, $E^2$ \cite{Schw,KKT,Zh2,Zh6} and, therefore, 
does {\em not} reduce to a standard 
eigenvalue problem. The physical problem is rather to find zero modes $\phi$ of ${\cal L}$.
Differential equations for the Rabi models are not even Fuchsian, having 
an irregular singular point at infinity. 
Obviously in analyzing a given 2nd order ODE with polynomial coefficients one can 
always switch from the above form into a Schr\"odinger equation (SE) 
(also known as normal) form, where the first derivative term has been 
eliminated and the coefficient of $d_z^2$ has been set to one \cite{KO,GKO}. 
Such a transformation leads to an energy dependent potential and to a non-Sturm-Liouville problem.

Inspired by $2$nd order ODE which occur 
when trying to solve the Rabi models \cite{Rb,Schw,TAP,TPG,AMdn,Zh2,Zh6,NLL,EB1},
we developed general necessary and sufficient conditions 
for the existence of a polynomial solution of $n$th degree of such equations.
In recent years, 
the corresponding differential operator ${\cal L}$
for a (driven) quantum Rabi model \cite{KKT,Zh2,Zh6},
the two-photon and two-mode quantum Rabi models \cite{Zh2,Zh6},
and the generalized Rabi model \cite{TAP} was shown 
to be expressible as a bilinear combination of $sl_2$ algebra generators, 
and hence to be an element of the enveloping algebra $\mathcal U(sl_2)$ for a certain 
choice of model parameters and of energy.
For a typical eigenvalue problem, ${\cal L}\psi=\lambda\psi$, one can always find 
$n+1=2j+1$ polynomial solutions 
corresponding to different eigenvalues $\lambda_l$ of ${\cal L}$ in a corresponding 
$sl_2$ module characterized by the spin $j$. The constant term $c_0$ of ${\cal L}$ is 
a free parameter that can always be absorbed to an eigenvalue $\lambda_l$.
However, in the case of the Rabi models, the value of $c_0$ is fixed and 
only zero modes $\phi$ of ${\cal L}$, which satisfy ${\cal L}\phi=0$, are physical solutions. 
One can easily show that the latter problem can have at most a single polynomial solution. 
We show that with ${\cal L}\in\mathcal U(sl_2)$, the ODE ${\cal L}\psi=0$ need not have in general any
polynomial solution. In particular, $sl_2$ algebra does not explain why the Juddian isolated exact 
solutions can be analytically computed \cite{Jd,Ks,Li15,Wa,KRW}, whereas the remaining part of the spectrum not. 
The Rabi models are thus an {\em unusual}
example of {\em quasi-exactly solvable} (QES) models \cite{KO,GKO,TU,Trb,ShT,Shfm,Trb1,Tams,FGR,KKT}. 
The QES models are distinguished by the fact 
that a {\em finite} number of their eigenvalues and 
corresponding eigenfunctions can be determined algebraically
\cite{KO,GKO,TU,Trb,ShT,Shfm,Trb1,Tams,FGR}. 
Initially, QES was essentially a synonym for $sl_2$ algebraization 
in one dimensional quantum mechanical problems \cite{KO,GKO,TU,Trb,ShT,Shfm,Trb1,Tams,FGR,KKT}, 
up to the point that no difference was made between the terms {\em Lie-algebraic}
and {\em quasi-exactly solvable} in the literature. 
The reason behind this was that $sl_2$ is the only algebra of
first order differential operators with finite dimensional invariant modules. 
{\em Burnside}'s classical theorem ensures that every
differential operator which leaves the space $\mathcal{P}_n(z)$ invariant
belongs to the enveloping algebra $\mathcal U(sl_2)$, since
$\mathcal P_n(z):={\rm span}\{1,z,\ldots,z^n\}$ is an irreducible module for the $sl_2$ action.

The article is organized as follows.
On defining the {\em grade} of a term $z^m d_z^l$ as integer $m-l$, Sec. \ref{sc:str} 
introduces other necessary definitions to perform a gradation slicing of a given
ordinary linear differential equation (ODE) with polynomial coefficients,
\begin{equation}
{\cal L} S_n(z)= \left \{A(z)\frac{d^2}{dz^2}+B(z)\frac{d}{dz}+C(z)\right\}S_n(z)=0,
\label{ODE}
\end{equation}
where
\begin{equation}
A(z)=\sum_{k=0} a_kz^k,~~~~B(z)=\sum_{k=0} b_kz^k,~~~~C(z)=\sum_{k=0} c_kz^k.
\label{yaoc}
\end{equation}
In Sec. \ref{sc:thrs} basic theorems are formulated that yield necessary and sufficient conditions 
for the existence of a polynomial solution of $n$th degree.
The necessary condition in general constraints energy in a parameter space for each given $n$ to a different
baseline [e.g. Eqs. (\ref{c+cn}), (\ref{Rabi-solution-E}), 
(\ref{2-photon-energy}), (\ref{2-mode-energy}), (\ref{gRnc}) below].
Provided that the highest grade of the terms $z^m d_z^l$ of ODE (\ref{ODE}) is $\gamma$,
there are $\gamma$ recursively defined constraints to be satisfied
for a polynomial solution on the $n$th baseline to exist.
It turns out that, with the exception of the generalized Rabi model \cite{TAP}, 
each of the Rabi models considered in this article is characterized by
an ordinary linear differential equation comprising terms with highest grade $\gamma$, 
the lowest grade $\gamma_*$, gradation width $w$, the highest grade
slice ${\cal F}_{\gamma}$, and its induced multiplicator $F_{\gamma}(n)$ as follows:
\begin{equation}
\gamma=1,\qquad \gamma_*=-2,\qquad w=4,\qquad 
{\cal F}_{\gamma}=b_2 z^2 d_z + c_1z,\qquad F_{\gamma}(n)=nb_2+c_1.
\label{grmch}
\end{equation}
For the generalized Rabi model one finds, depending on parameters, either $\gamma=1$ or $\gamma=2$.
The corresponding constraints are shown to (i) reproduce known constraint
polynomials for the usual and driven Rabi models \cite{Jd,Ks,Li15,Wa,KRW} (cf. Figs. \ref{fgkus5}, \ref{fggkus9}) 
and (ii) generate hitherto unknown constraint polynomials for the two-mode, 
two-photon, and the generalized Rabi models.
Usual road to constraint polynomials required to reveal an ingenious Ansatz 
for the polynomial solutions. For example, the original Kus construction \cite{Ks} consisted 
in an insightful observation
that an exact polynomial solution of the Rabi model on the $n$th baseline can be constructed as a finite
linear combination of the solutions $\Phi_l^\pm$ of a {\em displaced harmonic oscillator}
($\Delta=0$ limit of the Rabi model) from all baselines $l\le n$ and of the same parity. 
An  analogous approach was attempted later by Emary and Bishop \cite{EB1} in the case of two-photon Rabi model, 
and the others in the case of the driven Rabi model \cite{Li15,Wa,KRW}.
Yet the origin of constraint polynomials remained mysterious. It was not a priori clear if they at all exist.
In this regard Theorem 3 of Sec. \ref{sc:thrs} yields a recipe for determining
constraint polynomials by a downward recurrence (\ref{gm0rc}), (\ref{gmrc}) 
with well defined coefficients for any $\gamma>0$ problem, and in particular for 
any conceivable Rabi model generalization.
In Theorem 4 of Sec. \ref{sc:thrs} we have succeeded to generalize an important result of Zhang 
(cf. Eqs. (1.8-10) of Ref. \cite{Zh}) obtained for $2$nd order ODE's to the case of arbitrary $\gamma\ge 0$. 
A $sl_2$ algebraization with spin $j=n/2$ is shown in Sec. \ref{sc:gsl2c}  
to be equivalent merely to the necessary condition for the existence of a polynomial solution of $n$th degree.
A lemma is formulated which yields necessary condition for a 
spectral problem $T_2\psi=\lambda\psi$, where $\lambda\in\mathbb{C}$ is an eigenvalue,
to have degenerate energy levels in an invariant $sl_2$ module of spin $j$.

In Sec. \ref{sc:qrm} our approach is illustrated in detail on the example of 
the usual quantum Rabi model. A driven Rabi model is considered in section \ref{sc:drm}.
Nonlinear two-photon and two-mode Rabi models are dealt with in section \ref{sec:nlrm}, and the generalized Rabi model 
is the subject of section \ref{sec:grm}. In each of the above cases explicit expressions of 
the recurrence coefficients for the constraint polynomials are presented.
Our results open a number of different avenues of further research which 
are discussed in Sec. \ref{sec:disc}. We then conclude in Sec. \ref{sec:conc}.
For the sake or presentation, Appendix \ref{cs:slc}
provides an overview of the basics of $sl_2$ algebraization.
Relevant features of $2$nd order linear differential equation 
with all solutions being polynomials are summarized in Appendix \ref{sc:eqps}.
Singular points at spatial infinity are dealt with in Appendix \ref{sc:spsi}.
Some other alternative forms of $2$nd order ODE of Refs. \cite{Schw,KKT} 
for the Rabi model are examined in Appendix \ref{sec:alt}.

\section{Gradation slicing}
\label{sc:str}
%%%%%%%%%%%%%%%%%%%%%%%%%%%%%%%%%%%%%
The subset of ODE (\ref{ODE}) where $A(z), B(z), C(z)$ are {\em polynomial} 
of degree at most $4$, $3$, $2$, respectively,
covers (i) all QES models within the context 
of $sl_2$ [cf. Eqs. (\ref{sl2qes}), (\ref{p432sl2n})] \cite{KO,GKO,TU,Trb,ShT,Shfm,Trb1,Tams,FGR},
(ii) all Fuchsian 2nd order ODE \cite{In1}
and (iii) {\em non-Fuchsian} 2nd order ODE of the present article [such as Eq. (\ref{Rabi-diff1}) below,
which has an irregular singular point at infinity (see Appendix \ref{sc:spsi})].

For the purpose of looking for (monic) polynomial solutions, 
\begin{equation}
S_n(z)=\prod_{i=1}^n(z-z_i)=\sum_{k=0}^n a_{nk} z^k \qquad \qquad (a_{nn}\equiv 1),
\label{snex}
\end{equation}
of the ODE (\ref{ODE}), or in general of
\begin{equation}
{\cal L} S_n(z)= \left \{\sum_{l=3} X_l(z)\frac{d^l}{dz^l} 
   + A(z)\frac{d^2}{dz^2}+B(z)\frac{d}{dz}+C(z)\right\}S_n(z)=0,
\label{ODEg}
\end{equation}
where $X_l(z)$ are also polynomials,
it is expedient to rearrange it into a more convenient form.
Obviously for any polynomial $S_n(z)$ the image ${\cal L} S_n(z)$ is also a polynomial.
The basic idea is to characterize the terms of the operator ${\cal L}$ 
which contribute to the same polynomial degree in the image ${\cal L} S_n(z)$.
In what follows we call (a positive or negative) integer ${\mathfrak g}=m-l$ the {\em grade}
of the term $z^m d_z^l$. The grade describes a change of the degree of a monomial
$z^n$ under the action $z^m d_z^l$. This is similar to the grading (\ref{e15}) 
of $sl_2$ generators (\ref{sl2r}) employed by Turbiner \cite{Trb1,Tams}.

We introduce the concept of a {\em gradation slicing} of an ODE (\ref{ODEg}), which comprises
the following steps:
\begin{itemize}

\item [({\bf S1})] Consider a given differential equation as a linear 
combination of terms $\sim z^m d_z^l$ and determine the grade of each term.

\item [({\bf S2})] Rearrange all the terms of the ODE according to their grade.
The subset ${\cal F}_{\mathfrak g} (z^m d_z^l; m-l\equiv {\mathfrak g})$ 
of the ODE with an identical grade ${\mathfrak g}$ will be called a {\em slice}.
Hence the differential equation can be recast as
\begin{equation}
{\cal L} S_n(z)= 
\sum_{{\mathfrak g} ={\mathfrak g}_{min}}^{{\mathfrak g}_{max}} {\cal F}_{\mathfrak g} S_n(z)=0,
\label{ODEgs}
\end{equation}
where the sum runs over all grades ${\mathfrak g}$. In what follows we will use an abbreviation
$\gamma={\mathfrak g}_{max}$ for the highest grade and 
$\gamma_*={\mathfrak g}_{min}$ for the lowest grade.

\end{itemize}
%%%%%%%%%%%%%%%

\vspace{.1in}

\noindent
{\bf Definition:}
A decomposition of original ordinary linear differential equation (\ref{ODEg}) into 
(\ref{ODEgs}) will be called gradation slicing.
We call the {\em grade} of an ordinary linear differential equation the highest grade $\gamma$.
A {\em width} $w$ of the gradation slicing will be called the integer $w:=\gamma-\gamma_*+1$.
Define a function $F_{\mathfrak g}(k)$ by
\begin{equation}
{\cal F}_{\mathfrak g} z^k:=F_{\mathfrak g}(k) z^{k+{\mathfrak g}}.
\nonumber
\end{equation}
We shall call the function $F_{\mathfrak g}(k)$ an {\em induced multiplicator} 
corresponding to the slice ${\cal F}_{\mathfrak g}$.

\hfill $\Box$

\vspace{.1in}

The width counts the number of possible slices with the grade between the minimal 
and maximal grades, $\gamma_*$ and $\gamma$, respectively. 
Unless $C(z)\equiv 0$ is identically zero, one has always $\gamma\ge 0$.
In what follows, we shall assume that $\gamma_*\le 0$. The case $\gamma_*>0$ 
can always be reduced to the case 
$\gamma_*=0$ by factorizing $z^{\gamma_*}$ out of the polynomial coefficients 
of the differential equation (\ref{ODEg}).

\vspace{.1in}

\noindent
{\bf Remark 1}:
A hypergeometric equation is characterized by $\gamma= 0$, $\gamma_*=-2$, and $w=3$.
A typical Heine-Stieltjes problem \cite{Heine1878,Stl,Sz,Schhs,Shp,AMu,AGM}, where 
$A(z), B(z), C(z)$ are {\em polynomials} of exact degree $N+2$, $N+1$, $N$, respectively,
is grade $\gamma=N$, $\gamma_*=-2$, $w=N+3$ problem.

\vspace{.1in}

It turns out that each of the Rabi models considered in this article is described by
an ordinary linear differential equation characterized by 
$\gamma=1$, $\gamma_*=-2$, $w=4$, ${\cal F}_{\gamma}$, and induced multiplicator 
$F_{\gamma}(n)$ as summarized by (\ref{grmch}).
The lowest grade and width are not absolute invariants
of an ODE, because they may depend on the origin of coordinates.

Obviously the condition that the slice with the highest grade $\gamma$
annihilates a monomial of degree $n$, ${\cal F}_{\gamma} z^n=0$, 
provides a {\em necessary} condition for the existence 
of a polynomial solution of degree $n$,
\begin{equation}
F_{\gamma}(n) =0.
\label{gnc}
\end{equation}
Provided that $F_{\gamma}(n)$ depends on energy $E$ 
one can consider the condition (\ref{gnc}) as equation for $E$. 
(For the Rabi models considered here this will be typically a linear equation.)
Its solution $E=E(n)$ constraints energy $E$ in the parameter space [e.g. Eqs. (\ref{c+cn}),
(\ref{Rabi-solution-E}), (\ref{2-photon-energy}),
(\ref{2-mode-energy}), (\ref{gRnc}) below]. Therefore, we will refer to the condition $F_\gamma(n)=0$
also as the {\em baseline condition}, although it defines a line only in the case of the original 
Rabi model, where it depends on a single parameter $g$ \cite{Jd,Ks}.

For the ensuing analysis of Rabi models we need both necessary and sufficient conditions 
for the existence of a polynomial solution of $n$th degree. 
In what follows we shall distinguish two main alternative types of differential equations (\ref{ODEg}):
\begin{itemize}

\item[({\bf A1})] a {\em conventional} one when the highest derivative term [e.g. $A(z)d_z^2$ in (\ref{ODE})]
does contribute to the slice ${\cal F}_{\gamma}$ with the highest grade $\gamma$. Applied to (\ref{ODE}), 
the alternative occurs if
\begin{equation}
\deg B\le\deg A-1,\qquad \deg C\le\deg A-2,
\label{1altr}
\end{equation}
where strict equality applies in at least one of the above cases.

\item[({\bf A2})] an {\em anomalous} one, when when the highest derivative term [e.g. $A(z)d_z^2$ 
in (\ref{ODE})] does {\em not}
contribute to ${\cal F}_{\gamma}$. Applied to (\ref{ODE}), the alternative occurs if
\begin{equation}
\deg B\ge \deg A,\qquad \deg C=\deg B-1.
\label{2altr}
\end{equation}
\end{itemize}
%%%%%%%%%%%%%
An example of the alternative ({\bf A1}) are the Fuchsian equations, which include
a hypergeometric one, and the Heine-Stieltjes problem \cite{Heine1878,Stl,Sz,Schhs,Shp,AMu,AGM}.
The alternative ({\bf A2}) is usually omitted in the analysis 
of polynomial solutions. Yet for all Rabi model examples which follow, the alternative ({\bf A2}) 
will be the only relevant
one. Anomalous alternative ({\bf A2}) can be encountered also in other problems 
(cf. Eq. (5) of Ref. \cite{Tr94}; Eq. (31) of Ref. \cite{CH} for relative motion of two electrons 
in an external oscillator potential).
Obviously, one has automatically $B\ne -A'$ for the alternative ({\bf A2}).
Therefore, the necessary condition (\ref{vshc}) 
for a 2nd order ODE (\ref{ODE}) with fixed polynomial coefficients $A(z), B(z), C(z)$ to have two linearly 
independent (and hence to possess only) polynomial solutions is {\em always} violated.
Consequently if Eq. (\ref{ODE}) has a polynomial solution, such a polynomial solution 
is necessarily {\em unique}.

\section{Basic theorems}
\label{sc:thrs}
%%%%%%%%%%%%%%%%%%%%%%%%%%%%%%%%%%%%%
In this section general necessary and sufficient conditions 
for the existence of a polynomial solution of $n$th degree are formulated.

\subsection{General theory}
\label{sc:gthr}
%%%%%%%%%%%%%%%%%%%%%%%%%%%%%%%%%%%%%
The condition that $S_n(z)$ solves (\ref{ODEgs}) is equivalent to that all 
the coefficients of respective powers of $z$ of the image ${\cal L} S_n(z)$ of $S_n(z)$ vanish.
The latter brings us to the linear system of equations
\begin{align}
 F_{\gamma-1}(n)+ a_{n,n-1} F_{\gamma} (n-1)=0,&
\nonumber\\
 F_{\gamma-2}(n)+ a_{n,n-1} F_{\gamma-1} (n-1)+ a_{n,n-2} F_{\gamma} (n-2)=0,&
\nonumber\\
 \vdots\qquad \vdots \qquad \vdots\qquad \vdots \qquad \vdots\qquad \vdots\qquad 
 \vdots \qquad \vdots\qquad \vdots \qquad \vdots&
\nonumber\\
a_{n,\gamma+2-\gamma_*}
F_{\gamma_*}(\gamma+2-\gamma_*)+ \ldots + a_{n3} F_{\gamma-1} (3)+ a_{n2} F_{\gamma} (2)=0,&
\nonumber\\
a_{nw}
F_{\gamma_*}(w)+ \ldots + a_{n2} F_{\gamma-1} (2)+ a_{n1} F_{\gamma} (1)=0,&
\nonumber\\
a_{n,w-1}
F_{\gamma_*}(w-1)+ \ldots + a_{n1} F_{\gamma-1} (1)+ a_{n0} F_{\gamma} (0)=0,&
\label{gm0rc}
\end{align}
where each line summarizes all the terms contributing to the same power of $z$, beginning 
from $n-1+\gamma$ of the first equation down to $\gamma$ of the last equation. We recall that 
$w=\gamma-\gamma_*+1$ is the gradation slicing width.
If one tries to determine the coefficients $a_{nk}$ of $S_n(z)$ in the expansion (\ref{snex}) 
by direct substitution into underlying differential equation, the width $w$ thus yields the length of 
a downward recurrence. For $\gamma=0$ both $S_n(z)$ and its image ${\cal L} S_n(z)$ are 
polynomials of the same degree.
Moreover, one has necessarily $\gamma_*\le 0$. We have the following Theorem.

\vspace{.1in}

\noindent
{\bf Theorem 1:}
A necessary and sufficient conditions for the ODE (\ref{ODEg}) with the grade $\gamma=0$ 
to have a {\em unique} polynomial solution of $n$th degree is that 
\begin{equation}
F_0(n)=0,\qquad F_0(k)\ne 0, ~~~ 0\le k<n,
\label{th1nsc}
\end{equation}
where the second condition applies for $n\ge 1$.
 \hfill $\Box$

\vspace{.1in}

\noindent {\em Proof}:
The condition $F_0(n)=0$ is nothing but the baseline condition (\ref{gnc}) in the special case
$\gamma=0$, and is obviously necessary. In order to demonstrate sufficiency, 
note that the second condition $F_0(k)\ne 0$ ensures that each subsequent line in the system (\ref{gm0rc}) 
of $n$ equations, when progressing from the very top down, 
enables one to {\em uniquely} determine newly appearing coefficient (i.e. $a_{n,n-l}$ in the $l$th line) and thus
to determine at the end a unique set of coefficients $a_{nk}$, $0\le k<n$. The initial
condition $a_{nn}=1$ is used here to simply fix an arbitrary irrelevant multiplication factor. 
The point of crucial importance is that for (and only for) $\gamma=0$ 
the image ${\cal L} S_n(z)$ and $S_n(z)$ are polynomials of the same degree.
Hence on summing up the lines of the above system
one recovers the original system of $n$ equations for $n$ unknown coefficients
$a_{nk}$, $0\le k<n$. Indeed, because by definition, and on substituting $\gamma=0$,
\begin{equation}
{\cal F}_{\gamma_*+m}z^{-\gamma_* -m-l}=F_{\gamma_*+m}(-\gamma_* -m-l) \equiv 0,\qquad 0\le m\le w-1,~~ l> 0.
\label{crc}
\end{equation}
Thereby the theorem is proven. \hfill $\Box$

\vspace{.1in}

\noindent
{\bf Corollary 1:}
If we drop ``{\em unique}" in Theorem 1, then the condition (\ref{gnc})
is both necessary and sufficient condition for the existence of a polynomial 
solution of an ODE of grade zero.

\vspace{.1in}

\noindent {\em Proof}: Apply Theorem 1 to the smallest nonnegative zero of $F_0$.
\hfill $\Box$

\vspace{.1in}

\noindent
{\bf Corollary 2:}
If $F_{\gamma}(k)$ is a {\em linear} function 
of $k$, there is always at most a single unique polynomial solution, because
a linear function can have at most a single root.
\hfill $\Box$

\vspace{.1in}

\noindent
{\bf Remark 2}:
For the hypergeometric equation characterized by $\gamma= 0$ and $w=3$, the system of equations 
(\ref{gm0rc}) reduces to the three-term recurrence relation (TTRR) 
studied exhaustively by Lesky \cite{Lsk}. $F_0(k)$ is a quadratic function of $k$ and there are, in principle,
possible two linearly independent polynomial solutions, because quadratic function has in general two roots
(cf. Appendix \ref{sc:eqps}). 

\vspace{.1in}

\noindent
{\bf Remark 3}: In the case of the Heine-Stieltjes problem \cite{Heine1878,Stl,Sz,Schhs,Shp,AMu,AGM}, 
the usual condition $B=-A'$ for ODE (\ref{ODE}) to have only polynomial solutions 
(see Appendix \ref{sc:eqps} for more detail) requires that for some $n$
\begin{equation}
F_2(n)=n(n-1)a_N + nb_{N-1}+c_{N-2}=n(n-1-N)a_N+c_{N-2} =0.
\nonumber
\end{equation}
\hfill $\Box$

\vspace{.1in}

\noindent
{\bf Remark 4}:
Theorem 1 does not rely on, and is independent of, the Frobenius analysis of 
a regular singular point of $2$nd order ODE
(see for instance Chap. 10.3 of Whittaker and Watson \cite{WW}, or Chap. 5.3 of Hille \cite{Hi}).
Yet there are many parallels between the two approaches.
In Theorem 1 the condition $F_0(n)=0$ gives an entry point to a {\em downward} recurrence.
In the Frobenius analysis, one needs instead an entry point for an {\em upward} 
recurrence. Such an entry point is provided by 
the solutions of the so-called {\em indicial equation}, which can be 
viewed as $F_{\gamma^*}(s)=0$, $s\in\mathbb{C}$. Hence in the Frobenius analysis
one is instead of the highest grade slice ${\cal F}_{\gamma}$ concerned with
the lowest grade slice ${\cal F}_{\gamma^*}$.
If $s$ is a solution, i.e. $F_{\gamma^*}(s)=0$, then the condition $F_{\gamma^*}(n+s)\ne 0$, $n\ge 1$,
guarantees that a {\em unique}, in general infinite, set of coefficients of a solution
at a regular singular point can be determined by the relevant recurrence.
\hfill $\Box$

\vspace{.1in}

As explained above, we assume that $\gamma_*\le 0$,
which can always be achieved by a suitable factorization of the polynomial coefficients of 
differential equation (\ref{ODEg}).

\vspace{.1in}

\noindent
{\bf Theorem 2:}
A necessary and sufficient conditions for the ODE (\ref{ODEg}) with the grade $\gamma>0$ 
to have a unique polynomial solution is that, in addition to the conditions (\ref{th1nsc}) which determine 
the unique set of coefficients $\{a_{nk}\}_{k=0}^n$ by the recurrence (\ref{gm0rc}) of Theorem 1,
the subset $\{a_{n0},a_{n1},\ldots,a_{n,w-2}\}$ of the coefficients $\{a_{nk}\}_{k=0}^n$
satisfies additional $\gamma$ constraints:
\begin{align}
P_\gamma:=a_{n,w-2}
F_{\gamma_*}(w-2)+ \ldots + a_{n1} F_{\gamma-2} (1)+ a_{n0} F_{\gamma-1} (0)=0,&
\nonumber\\
 \vdots\qquad \vdots \qquad \vdots\qquad \vdots \qquad \vdots\qquad 
\vdots\qquad \vdots \qquad \vdots\qquad \vdots \qquad \vdots&
\nonumber\\
P_2:= a_{n,1-\gamma_*} F_{\gamma_*} (1-\gamma_*)+\ldots + a_{n1} F_0 (1)+ a_{n0} F_{1} (0)=0,&
\nonumber\\
P_1:=  a_{n,-\gamma_*} F_{\gamma_*} (-\gamma_*)+\ldots + a_{n1} F_{-1}(1)+ a_{n0} F_0 (0)=0.&
\label{gmrc}
\end{align}

\vspace{.1in}

\noindent {\em Proof}:
According to the definition, ${\cal F}_{\gamma} z^0 \sim z^{\gamma}$. 
Therefore, whenever $\gamma>0$, the recurrence (\ref{gm0rc}) does not take into account 
the terms $\sim z^k$ of degree $k<\gamma$ of the image ${\cal L} S_n(z)$. 
There are exactly $\gamma$ of such polynomial terms with $k=0,\ldots,\gamma-1$.
One can verify that the vanishing of the coefficients of $z^k$, $0\le k<\gamma$ 
amounts to solving the system (\ref{gmrc}).
The vanishing of the coefficients of $z^k$, $k<\gamma$ thus imposes 
$\gamma$ constraints on the (up to a multiplication by a constant) unique set of coefficients $a_{nk}$.
\hfill $\Box$

\vspace{.1in}

\noindent
{\bf Remark 5}:
Grade $\gamma< 0$ problem has always a polynomial
solution, because it leads to a system of $(n-|\gamma|)< n$ equations for $n$ unknowns.
Obviously, whenever $\gamma>0$, the differential operator ${\cal L}$ is {\em not} 
exactly solvable. The image ${\cal L} S_n(z)$ of $S_n(z)$ is a polynomial 
of $(n+\gamma)$th degree. The vanishing of all the polynomial coefficients
of the image ${\cal L} S_n(z)$ then imposes $(n+\gamma)$ different conditions.
\hfill $\Box$

\vspace{.1in}

\noindent
{\bf Remark 6}: 
After imposing on the energy $E$ one of the baseline conditions $F_\gamma(n)=0$, 
the energy is expressed as a function of model parameters 
[e.g. Eqs. (\ref{c+cn}), (\ref{Rabi-solution-E}), (\ref{2-photon-energy}),
(\ref{2-mode-energy}), (\ref{gRnc}) below]. Therefore after imposing the baseline condition,
the coefficients $F_{\mathfrak g}(k)$ of the recurrence (\ref{gm0rc}) and of (\ref{gmrc}) cease 
to depend on $E$. When solving for the expansion coefficients $a_{nk}$ by the $w$-term 
{\em downward} recurrence (\ref{gm0rc}), each subsequent $a_{nk}$, beginning from $k=n-1$ 
in the first equation down to $k=0$ in the last equation,
is obtained by dividing the corresponding equation line by $F_\gamma(k)$. 
The necessary and sufficient conditions for the ODE (\ref{ODEg}) with the grade $\gamma>0$ 
to have a unique polynomial solution ensures that the product $\prod_{k=n-1}^0 F_\gamma(k)\ne 0$.
Provided that the coefficients $F_{\mathfrak g}(k)$ are {\em polynomials} in model parameters
[e.g. examples (\ref{fffr}), (\ref{fffrd}), (\ref{fffr2p}), (\ref{fffr2m}), 
(\ref{fffrgR}) below], each $P_{\mathfrak g}$, ${\mathfrak g}=1,\ldots,\gamma$ defined by Eq. (\ref{gmrc}), 
when multiplied by the product $\prod_{k=n-1}^0 F_\gamma(k)\ne 0$, is 
necessarily a {\em polynomial} in model parameters. The resulting constraint polynomials are  
defined by the $w$-term recurrence (\ref{gm0rc}), which yields the 
unique solution for $a_{n2}$, $a_{n1}$, $a_{n0}$, that is substituted into 
the constraints (\ref{gmrc}) and each of the constraints (\ref{gmrc}) 
is multiplied by $\prod_{k=n-1}^0 F_\gamma(k)\ne 0$,
\hfill $\Box$

\vspace{.1in}

\noindent We have thus proven the following fundamental result:

\vspace{.1in}

\noindent
{\bf Theorem 3:}
Provided that each $F_{\mathfrak g}(k)$ in (\ref{gm0rc}) and (\ref{gmrc})
is a polynomial in model parameter(s) and the hypotheses of Theorem 2 are satisfied, 
each recursively determined $P_{\mathfrak g}$, ${\mathfrak g}=1,\ldots,\gamma$ 
of Eq. (\ref{gmrc}) is proportional to a polynomial in model parameter(s).
\hfill $\Box$

\vspace{.1in}

\noindent
{\bf Remark 7}:
The Rabi models in this article are all characterized by the same grading parameters
summarized by (\ref{grmch}). Given that $w=4$, the recurrence system (\ref{gm0rc}), (\ref{gmrc}) reduces 
to a {\em downward} four-term recurrence relation (FTRR) for the coefficients $a_{nk}$, $k<n$,
of $S_n(z)$. The necessary condition (\ref{gnc}) for the existence of a polynomial solution
becomes in view of (\ref{grmch})
\begin{equation}
nb_2+c_1=0,\qquad n\ge 0.
\label{c+cgn}
\end{equation}
For $\gamma=1$ there is a single constraint (\ref{gmrc}) to be satisfied,
\begin{equation}
P_1:= a_{n2} F_{-2}(2)+ a_{n1} F_{-1} (1)+ a_{n0} F_0 (0)=0,
\label{rmgmrc}
\end{equation}
to guarantee the existence of a unique polynomial solution.
For the Rabi models considered here all the coefficients $F_{\mathfrak g}(k)$ are polynomials in physical parameters
such as the coupling strength $g$, detuning $\Delta$, and frequency $\omega$ in Eq. (\ref{rabih})
[e.g. examples (\ref{fffr}), (\ref{fffrd}), (\ref{fffr2p}), (\ref{fffr2m}), 
(\ref{gRnc}) below]. 
\hfill $\Box$

\vspace{.1in}

\noindent
{\bf Corollary 3:} For the Rabi model (\ref{rabih}) the constraint (\ref{rmgmrc}) for each baseline is 
equivalent to the corresponding {\em Kus} polynomial \cite{Ks}. For the driven Rabi model, 
the constraint (\ref{rmgmrc}) on a given baseline is 
equivalent to the corresponding {\em generalized Kus} polynomial \cite{Li15,Wa,KRW}. 

\vspace{.1in}

The polynomial has to be equivalent to either the Kus polynomial \cite{Ks}, or 
generalized Kus polynomial \cite{Li15,Wa,KRW}, respectively,
because they express the necessary and sufficient conditions for the existence of 
a unique polynomial solution.
The equivalence will be illustrated on examples and numerically in 
Secs. \ref{sc:qrm}-\ref{sc:drm}.
\hfill $\Box$
\vspace{.1in}

\noindent
{\bf Remark 8}: 
For the grade $\gamma=2$ there would be two constraint polynomials. 
Common zeros of two different polynomials are the zeros of the so-called {\em resultant} \cite{MS}.
This is the case of the generalized Rabi model discussed in Sec. \ref{sec:grm} below.
\hfill $\Box$
\vspace{.1in}

\noindent
{\bf Theorem 4:}
Let the $2$nd order ODE (\ref{ODE}) be of grade $\gamma\ge 0$.
General {\em necessary} conditions on the coefficients of
the polynomial $C(z)$ for the ODE (\ref{ODE}) 
to have a polynomial solution $S_n(z)$ of degree $n$ with all zeros $z_i$ simple are
\begin{eqnarray}
c_{\gamma} &=& -n(n-1)a_{\gamma+2}-nb_{\gamma+1},
\label{c2-expression}\\
c_{\gamma-1} &=& -\left[2(n-1)a_{\gamma+2}+b_{\gamma+1}\right]\sum_{i=1}^nz_i-n(n-1)a_{\gamma+1}-nb_{\gamma},
\label{c1-expression}
\\
c_{\gamma-2} &=& -\left[2(n-1)a_{\gamma+2}+b_{\gamma+1}\right]\sum_{i=1}^nz_i^2-2a_{\gamma+2}\sum_{i<l}^nz_iz_l
\nonumber\\
 & &-\left[2(n-1)a_{\gamma+1}+b_{\gamma}\right]\sum_{i=1}^nz_i-n(n-1)a_{\gamma}-nb_{\gamma-1}.
\label{c0-expression}
\end{eqnarray}
\hfill $\Box$

\vspace{.1in}

\noindent Note in passing that if a zero $z_i$ of $S_n(z)$ were not simple, then the Wronskian of $S_n(z)$
with any other (not necessary polynomial) nonsingular function would be 
zero at any multiple root $z_i$ of $S_n(z)$
(see Appendix \ref{sc:eqps} for more detail).
Here and below the coefficients $a_k, b_k, c_k$ with a negative subscript are assumed to
be identically zero.

\vspace{.1in}

\noindent {\em Proof}:
According to the hypothesis we have
\begin{eqnarray}
{\cal F}_{\gamma} &=& a_{\gamma+2} z^{\gamma+2}d_z^2 
   +b_{\gamma+1} z^{\gamma+1}d_z +c_{\gamma}z^{\gamma},
\nonumber\\
{\cal F}_{\gamma-1} &=& a_{\gamma+1} z^{\gamma+1}d_z^2
   +b_{\gamma} z^{\gamma}d_z +c_{\gamma-1} z^{\gamma-1},
\nonumber\\
{\cal F}_{\gamma-2} &=& a_{\gamma} z^{\gamma}d_z^2 
    +b_{\gamma-1} z^{\gamma-1}d_z +c_{\gamma-2} z^{\gamma-2}.
\nonumber
\end{eqnarray}
In particular $F_{\gamma}(n) = n(n-1) a_{\gamma+2}  + n b_{\gamma+1} +c_{\gamma}$, and
he {\em necessary} condition (\ref{gnc}) for the existence 
of a polynomial solution of degree $n$ constraints the coefficient $c_{\gamma}$ immediately to
(\ref{c2-expression}).

The first of the recurrences of the system (\ref{gm0rc}) requires
\begin{equation}
n(n-1)a_{\gamma+1} + nb_{\gamma}+c_{\gamma-1} + a_{n,n-1} 
\left[
(n-1)(n-2)a_{\gamma+2} +(n-1) b_{\gamma+1} +c_{\gamma}\right]=0.
\label{annm1}
\end{equation}
On substituting for $c_{\gamma}$ from (\ref{c2-expression}) and solving
for $c_{\gamma-1}$ yields
\begin{equation}
c_{\gamma-1} = 
\left[2(n-1)a_{\gamma+2}+b_{\gamma+1}\right]a_{n,n-1} -n(n-1)a_{\gamma+1}-nb_{\gamma}.
\label{gncfc1}
\end{equation}
On taking into account that $a_{n,n-1}=-\sum_i z_i$, this leads immediately to
(\ref{c1-expression}).

We now continue with the second of the recurrences of the system (\ref{gm0rc}),
\begin{equation}
F_{\gamma-2}(n)
 + a_{n,n-1} F_{\gamma-1}(n-1)+ a_{n,n-2} F_{\gamma} (n-2)=0.
\label{gm0rc2}
\end{equation}
One has
\begin{eqnarray}
F_{\gamma-2}(n) &=& n(n-1) a_{\gamma} +n b_{\gamma-1} +c_{\gamma-2},
\nonumber\\
F_{\gamma-1}(n-1) &=& (n-1)(n-2) a_{\gamma+1} +(n-1) b_{\gamma} +c_{\gamma-1}
\nonumber\\
 &=& 
\left[2(n-1)a_{\gamma+2}+b_{\gamma+1}\right]a_{n,n-1} +(n-1)[n-2-n]a_{\gamma+1}
 +[n-1 -n]b_{\gamma}
\nonumber\\
 &=& 
\left[2(n-1)a_{\gamma+2}+b_{\gamma+1}\right]a_{n,n-1} -2(n-1)a_{\gamma+1}
 - b_{\gamma},
\nonumber\\
F_{\gamma}(n-2) &=& (n-2)(n-3) a_{\gamma+2} +(n-2) b_{\gamma+1} +c_{\gamma}
\nonumber\\
 &=& [(n-2)(n-3)-n(n-1)] a_{\gamma+2} -2 b_{\gamma+1}
\nonumber\\
 &=& -2[2(n-1) a_{\gamma+2} + b_{\gamma+1}]+2 a_{\gamma+2},
\nonumber
\end{eqnarray}
where (\ref{c2-expression}) and (\ref{gncfc1}) were used in arriving at the
final results.
On substituting the above expressions back into (\ref{gm0rc2}) one arrives at 
\begin{align}
&
n(n-1) a_{\gamma} +n b_{\gamma-1} +c_{\gamma-2}
\nonumber\\
 & 
+ a_{n,n-1}\left\{
\left[2(n-1)a_{\gamma+2}+b_{\gamma+1}\right]a_{n,n-1} -2(n-1)a_{\gamma+1}
 - b_{\gamma}\right\}
\nonumber\\
 & - 2 a_{n,n-2} \left[2(n-1) a_{\gamma+2} + b_{\gamma+1}\right]+2 a_{n,n-2}a_{\gamma+2} =0.
\nonumber
\end{align}
Solving for $c_{\gamma-2}$ yields
\begin{align}
&
c_{\gamma-2} = \left[2(n-1)a_{\gamma+2}+b_{\gamma+1}\right] (2 a_{n,n-2} -a_{n,n-1}^2)
- 2a_{\gamma+2} a_{n,n-2} 
\nonumber\\
 & 
+ \left[2(n-1)a_{\gamma+1} + b_{\gamma}\right]a_{n,n-1} - n(n-1) a_{\gamma} - n b_{\gamma-1}.
\nonumber
\end{align}
To this end one makes use of
\begin{eqnarray}
&a_{n,n-1} = -\sum_i z_i,\qquad a_{n,n-2} = \sum_{i<l}^n z_iz_l, &
\nonumber\\
& a_{n,n-1}^2 = \left(\sum_{i=1}^n z_i\right)^2= \sum_{i=1}^n z_i^2 + 2\sum_{i<l}^n z_iz_l.&
\nonumber
\end{eqnarray}
Thereby one recovers (\ref{c0-expression}).
\hfill $\Box$

\vspace{.1in}

\noindent
{\bf Remark 9}:
In the special case of $\gamma=2$, the conditions (\ref{c2-expression})-(\ref{c0-expression}) reduce 
to those of Theorem 1.1 of Zhang (cf. Eqs. (1.8-10) of Ref. \cite{Zh}), where they were derived by means
of a functional Bethe Ansatz. Yet in the latter case the level of complexity
increases significantly with $\gamma$ and each $\gamma$-case has to be treated separately.
In contrast to that, the gradation slicing approach enables one to prove
the formulas of Theorem 4 for the coefficients $c_{\gamma-l}$, $l=0,1,2$, in one go.
Theorem 4 applies to both alternatives (\ref{1altr}) and (\ref{2altr}).
Note in passing that if $a_{\gamma+2}=b_{\gamma+1}=0$, the condition (\ref{c1-expression})
for $c_{\gamma-1}$ reduces to the necessary condition (\ref{c+cn}) for the existence 
of a polynomial solution of degree $n$.
Similarly for the condition (\ref{c0-expression}) for $c_{\gamma-2}$,
provided that additionally $a_{\gamma+1}=b_{\gamma}=0$.

\vspace{.1in}

\noindent
{\bf Remark 10}: Provided that the $n$ simple roots $z_i$ are required to satisfy the 
set of the {\em Bethe Ansatz} algebraic equations (cf. Eqs. (1.11) and (2.5) of Ref. \cite{Zh}),
\begin{eqnarray}
\lefteqn{
\sum_{l\neq i}^n\frac{2}{z_i-z_l}
+\frac{B(z_i)}{A(z_i)} =
}
\nonumber\\
&&
\sum_{l\neq i}^n\frac{2}{z_i-z_l}
+\frac{b_3z_i^3+b_2z_i^2+b_1z_i+b_0}{a_4 z_i^4+a_3z_i^3+a_2z_i^2+a_1z_i+a_0}=0,
~~~~i=1,2,\cdots,n,
\label{BAEs-general}
\end{eqnarray}
then the conditions together with those of Theorem 4 provide necessary and sufficient 
conditions for the coefficients of $C(z)$ of the ODE (\ref{ODEg}) 
with grade $\gamma=2$ to have a polynomial solution of $n$th degree with zeros $z_i$. Yet this does not
answer the question under which conditions has the system of the {\em Bethe Ansatz} algebraic equations a solution.
Theorem 1.1 of Zhang \cite{Zh} is rather a set of general compatibility conditions between the polynomial zeros $z_i$
that satisfy the {\em Bethe Ansatz} algebraic equations for a given $A(z)$ and $B(z)$ on one hand,
and the coefficients of $C(z)$ of the ODE (\ref{ODEg}) on the other hand.
Similarly to the Kus recipe \cite{Ks}, the Bethe Ansatz equations for a polynomial of $n$th degree 
have a solution only for a discrete set of model parameters, 
which corresponds to zeros of a certain polynomial in model parameters \cite{TAP}.

\subsection{$sl_2$ algebraization}
\label{sc:gsl2c}
%%%%%%%%%%%%%%%%%%%%%%%%%%%%%%%%%%%%%%%%%%%%%%%%%%%%%%%%%%%%%%%
The condition of a $sl_2$ algebraization is that a corresponding 
differential operator ${\cal L}$ can be expressed as a {\em normally ordered} bilinear 
combination of the $sl_2$ generators $J_\pm,\, J_0$,
\begin{eqnarray}
T_2 &=& C_{++} J_+ J_+ + C_{+0} J_+ J_0 + C_{00} J_0 J_0 + C_{0-} J_0 J_- 
+ C_{--} J_- J_- 
\nonumber\\
&& 
 +\, C_+ J_+ + C_0 J_0 + C_- J_- + C_*,
\label{sl2qes}
\end{eqnarray}
where $C_{\alpha \beta}, C_{\alpha}, C_* \in \mathbb{R}$ \cite{KO,GKO,TU,Trb,ShT,Shfm,Trb1,Tams,FGR}. 
Strictly speaking $T_2$ belongs to the central extension of $sl_2$ (cf. Theorem 2 of Ref. \cite{GKO}).
With a slight abuse of notation we continue writing $T_2\in\mathcal U(sl_2)$.
The properties of the $sl_2$ generators $J_\pm,\, J_0$ are summarized in Appendix \ref{cs:slc}.
 
Let us consider an anomalous ${\cal L}$ characterized by the grading parameters as in Eq. (\ref{grmch}). 
In order that such a ${\cal L}$ reproduces $T_2$ in Eq. (\ref{nosl2}), one has to have 
$\deg P_4=\deg P_3=2$ and $\deg P_2=1$ in Eq. (\ref{p432sl2n}).
The latter immediately requires $C_{++}=C_{+0} \equiv 0$ in Eqs. (\ref{sl2qes}), (\ref{p432sl2n}),
whereas $C_+$ there has to satisfy
\begin{equation}
C_+=b_2~~~~\mbox{and}~~~~ -2j C_+=c_1.
\label{c+c}
\end{equation}
The above two conditions on $C_+$ require that the coefficients $b_2$ and $c_1$ of 
the terms of the highest grade satisfy
\begin{equation}
2j b_2+c_1=0.
\label{sl2ac}
\end{equation}
The necessary condition (\ref{sl2ac}) for $sl_2$ algebraization reproduces the necessary condition (\ref{c+cgn})
for the existence of a polynomial solution of $n$th degree if and only if the spin $j$
of an irreducible $sl_2$ representation satisfies $2j=n$. 

There are at most $n$ different polynomial solutions of $n$th degree on each $n$th baseline \cite{Ks}.
The latter would not be surprising if we had an identical $T_2$ along whole given $n$th baseline.
Yet each of those polynomial solutions corresponds to a different $T_2\in\mathcal U(sl_2)$ 
(e.g. because $C_+$ depends on $g$ and the polynomial solutions are nondegenerate on a 
given (base)line in the $(E,g)$-plane \cite{Schw,Ks,AMops,AMtb,AMef}).

The remaining conditions for the $sl_2$ algebraization 
in the anomalous $\deg P_4=\deg P_3=2$ and $\deg P_2=1$ case are
\begin{eqnarray}
C_{00} &=& a_2,\qquad C_0=b_1+ (2j-1)a_2,
\nonumber\\ 
C_* &=& c_0 - j^2 C_{00} + jC_0= c_0 + jb_1 +j(j-1) a_2.
\label{sl2acb}
\end{eqnarray}
Because $C_*$ is an arbitrary constant in Eq. (\ref{sl2qes}), the conditions (\ref{sl2acb}) 
can be always satisfied. Therefore condition (\ref{sl2ac}) is both necessary and sufficient condition 
for $sl_2$ algebraization.

From the general necessary condition (\ref{c1-expression})
for the existence of a polynomial solution of $n$th degree of Theorem 4 we know 
that for an anomalous ${\cal L}$ characterized by the grading parameters as in Eq. (\ref{grmch}), i.e. with
$a_{\gamma+2}=a_3=0$, we have to have
\begin{equation}
c_0 = - b_2 \sum_{i=1}^nz_i-n(n-1)a_2-nb_1.
\label{c1-expressions}
\end{equation}
The general necessary condition (\ref{c1-expressions})
for the existence of a polynomial solution of $n$th degree of Theorem 4 for $2j=n$ can be then recast
in terms of $C_*$,
\begin{equation}
C_*=- b_2 \sum_{i=1}^nz_i-j(3j-1)a_2-jb_1.
\label{Cfrm}
\end{equation}

\subsubsection{Nondegenerate energy levels}
\label{cs:sldg}
%%%%%%%%%%%%%%%%%%%%%%%%%%%%%%%%%%%%%%%%%%%
Let us examine the conditions under which $T_2$ cannot have {\em degenerate}
energy levels in an invariant $sl_2$ module. A degeneracy can only occur if
the necessary condition (\ref{vshc}), $B=-A'$, is satisfied. 
The latter requires 
\begin{equation}
b_3+4a_4=0 \Longleftrightarrow (4j-3)b_3=0,
\nonumber
\end{equation}
which is impossible to satisfy for $b_3, a_4\ne 0$. Hence 
\begin{equation}
b_3=a_4\equiv 0 \Longleftrightarrow C_{++}=0.
\label{dc1}
\end{equation}
If $b_2, a_3\ne 0$, the condition (\ref{vshc}) requires 
\begin{equation}
b_2+3a_3=0 \Longleftrightarrow -(3j-4)C_{+0}+C_+ =0,
\nonumber
\end{equation}
or the constraint
\begin{equation}
C_+=(3j-4)C_{+0}.
\label{dc2}
\end{equation}
If $b_1, a_2\ne 0$, the condition (\ref{vshc}) requires 
\begin{equation}
b_1+2a_2=0 \Longleftrightarrow - (2j-3)C_{00} +C_0=0,
\nonumber
\end{equation}
or the constraint
\begin{equation}
C_0=(2j-3)C_{00}.
\label{dc3}
\end{equation}
Eventually,
\begin{equation}
b_0+a_1=0 \Longleftrightarrow C_- - (j-1) C_{0-} =0.
\label{dc4}
\end{equation}
We have thus proven the following result:

\vspace{.1in}

\noindent
{\bf Lemma:}
If any of the conditions (\ref{dc1})-(\ref{dc4}) is violated, the 
spectral problem $T_2\psi=\lambda\psi$, where $\lambda\in\mathbb{C}$ is an eigenvalue,
{\em cannot} have degenerate energy levels in an invariant $sl_2$ module of spin $j$. 
\hfill $\Box$

\vspace{.1in}
We remind here that such a nondegeneracy is, as shown by Lesky \cite{Lsk}), 
an {\em intrinsic} feature of any 
2nd order ODE spectral problem defining an orthogonal polynomials system (OPS) \cite{Sz,Chi}.

Usual $sl_2$ algebraization means that the 
corresponding spectral problem possesses $(n+1)$ eigenfunctions in the 
form of polynomials of degree $n=2j$ for any given irreducible representation
of $sl_2$ of spin $j$. The foregoing analysis of anomalous grade $\gamma=1$ problems 
implies for the Rabi model problems that any given irreducible representation 
of $sl_2$ of spin $j$ can in the most optimal case add 
only at most a {\em single} new polynomial eigenfunction of degree $n=2j$ relative to a 
lower dimensional irreducible representation of $sl_2$ of spin $j-(1/2)$ - 
a kind of {\em onion} algebraization. The forthcoming examples will demonstrate
that more often than not no new polynomial eigenfunction will be added to the spectrum.

\section{The Rabi model}
\label{sc:qrm}
%%%%%%%%%%%%%%%%%%%%%%%%%%%%%%
For a theoretical investigation of the Rabi model it is expedient to 
work in an equivalent {\em single-mode spin-boson} picture,
which amounts to interchanging $\sigma_1$ and $\sigma_3$ in (\ref{rabih}).
The latter is realized by unitary transformation $\hat{H}_R=U_{13} \hat{H}_R U_{13}$,
where $U_{13}=(\sigma_1 + \sigma_3)/\sqrt{2}$.
Assuming $\hbar=1$, one arrives at
\begin{equation}
\hat{H}_R=\omega a^\dagger a+\Delta\sigma_z+g\,\sigma_x\left(a^\dagger+a\right),
\label{RabiH}
\end{equation}
where $\sigma_z=\sigma_1$, and $\sigma_x=\sigma_3$ becomes {\em diagonal}. 
The $2\times 2$ matrix $\hat{H}_R$ possesses the parity symmetry $\hat{\Pi}_{FG}=\op{R}\sigma_1$,
where unitary $\op{R}=e^{{\rm i}\pi \hat{a}^\dagger \hat{a}}$
induces {\em reflections} of the annihilation and creation operators:
$\hat{a}\rightarrow-\hat{a}$, $\hat{a}^\dagger\rightarrow-\hat{a}^\dagger$, 
and leaves the boson number operator $\hat{a}^\dagger \hat{a}$ invariant \cite{Schw,AMdn,Ks,AMep}.
$\hat{H}_R$ is immediately recognized to be of the Fulton-Gouterman form \cite{AMdn,FG}, 
where the Fulton-Gouterman symmetry operation is realized by $\hat{\Pi}_{FG}$.
The projected parity eigenstates $\Phi^\pm$ have generically two independent components.
The advantage of the Fulton and Gouterman form is that 
the projected parity eigenstates $\Phi^\pm$ are characterized by a {\em single}
independent component. 
In the Bargmann realization \cite{Brg}: $a^\dagger\rightarrow z,~a\rightarrow d/dz$, 
and the Hamiltonian $\hat{H}_R$ of Eq. (\ref{RabiH}) becomes a matrix differential operator.
After $\hat{H}_R$ is diagonalized in the spin subspace, 
the corresponding one-dimensional differential operators are found to be of Dunkl type \cite{AMdn}.
The Fulton-Gouterman form and the one-dimensional differential operators of Dunkl type 
can also be determined for all the remaining Rabi models discussed here \cite{AMdn}.

In terms of the two-component wave function 
$\psi(z)=\left(\begin{smallmatrix}
\psi_+(z)\\
\psi_-(z)
\end{smallmatrix} \right)$,
the time-independent Schr\"odinger equation gives rise to a coupled system of 
two first-order differential equations (cf. Eqs. (2.4a-b) of Kus \cite{Ks}) 
\begin{eqnarray}
&&(\omega z+g)\frac{d}{dz}\, \psi_+(z)=(E-gz)\psi_+(z)-\Delta\psi_-(z),
\nonumber\\
&&(\omega z-g)\frac{d}{dz}\, \psi_-(z)=(E+gz)\psi_-(z)-\Delta\psi_+(z).
\label{Rb1sto}
\end{eqnarray}
If $\Delta=0$ these two equations decouple and reduce to the differential 
equations of two uncoupled
displaced harmonic oscillators which can be exactly solved separately \cite{Schw,Zh2}. 
For this reason we will concentrate on the $\Delta\neq 0$ case in the following.

With the substitution 
\begin{equation}
\psi_\pm(z)=e^{-gz/\omega}\phi_\pm(z),
\nonumber %%\label{psisb}
\end{equation}
and on eliminating $\phi_-(z)$ from the system we obtain the uncoupled 
$2$nd order differential equation for $\phi_+(z)$ \cite{Zh2,Zh6},
\begin{eqnarray}
&&\left[(\omega z+g)\frac{d}{dz}
-\left(\frac{g^2}{\omega}+E\right)\right]\phi_+(z)=-\Delta\phi_-(z),
\nonumber\\
&&\left[(\omega z-g)\frac{d}{dz}
-\left(2gz-\frac{g^2}{\omega}+E\right)\right]\phi_-(z)=-\Delta\phi_+(z).
 \label{Rabi-diff}
\end{eqnarray}
(The substitution $\psi_\pm(z)=e^{gz/\omega}\tilde\phi_\pm(z)$ merely interchanges 
the roles of $\phi_\pm$ and does not bring anything new.)

Eliminating $\phi_-(z)$ from the system we obtain the uncoupled differential 
equation for $\phi_+(z)$,
\begin{eqnarray}
\left[(\omega z-g)\frac{d}{dz}-\left(2gz-\frac{g^2}{\omega}+E\right)\right]
 \left[(\omega z+g)\frac{d}{dz}-\left(\frac{g^2}{\omega}+E\right)\right]\phi_+=\Delta^2\phi_+.
\nonumber
\end{eqnarray}
Explicitly, 
\begin{eqnarray}
&&(\omega z-g)(\omega z+g)\frac{d^2\phi_+}{dz^2}
+\left[-2\omega gz^2+(\omega^2-2g^2-2E\omega)z
 +\frac{g}{\omega}(2g^2-\omega^2)\right]\frac{d\phi_+}{dz}
\nonumber\\
&&~~~~~~~~~~~~~~~~~~
+\left[2g\left(\frac{g^2}{\omega}+E\right)z+E^2-\Delta^2-\frac{g^4}{\omega^2}
 \right]\phi_+\equiv {\cal L}\phi_+=0.
\label{Rabi-diff1}
\end{eqnarray}
Because the polynomial coefficients $b_1, c_1, c_0$ are all energy dependent,
this is obviously {\em not} a standard eigenvalue problem.
As announced earlier in Eq. (\ref{grmch}), the grading characteristics of 
the differential equation are $\gamma=1$, $w=4$, and
\begin{eqnarray}
&{\cal F}_1 = b_2d_z+c_1=-2\omega gz^2d_z +2g\left(\frac{g^2}{\omega}+E\right)z,&
\nonumber\\
&F_1(n) = -2n \omega g+2g\left(\frac{g^2}{\omega}+E\right),&
\nonumber\\
&{\cal F}_0 = \omega^2 z^2 d_z^2+ (\omega^2-2g^2-2E\omega) z d_z+ E^2-\Delta^2-\frac{g^4}{\omega^2},
\nonumber\\
&{\cal F}_{-1} = \frac{g}{\omega}(2g^2-\omega^2) d_z,\qquad  {\cal F}_{-2} =-g^2d_z^2.&
\nonumber
\end{eqnarray}
$F_1(n)$ is a {\em linear} function of $n$ and, according to Corollary 2, only 
one of any two linearly independent solution can by a polynomial.
One can arrive at the same conclusion by noting independently that,
in the case of Eq. (\ref{Rabi-diff1}), we have $\deg A=\deg B=2$, $\deg C=1$,
and the anomalous alternative ({\bf A2}) applies.
Hence any polynomial solution of Eq. (\ref{Rabi-diff1}) 
is necessarily {\em unique} (assuming $A(z), B(z), C(z)$ being
{\em fixed} polynomials). The other (nonpolynomial) solution is not normalizable
(e.g. does not belong to a corresponding Bargmann space \cite{Brg}), which 
follows from {\em non-degeneracy} of spectrum in each of the even and odd 
parity subspaces \cite{AMops,AMtb,AMef}.

One can verify that the necessary condition (\ref{gnc}) for 
the existence of a polynomial solution of degree $n$, $F_1(n)=0$, reduces to (\ref{c+cgn}) and leads to
\begin{equation}
E_n=n\omega -\frac{g^2}{\omega}\cdot
\label{c+cn}
\end{equation}
This is the familiar {\em baseline} condition for the Rabi model \cite{Schw,Ks}.
Each baseline corresponds to an exact energy level of a {\em displaced harmonic 
oscillator}, which is the $\Delta=0$ limit of the Rabi model \cite{Schw}.
The baseline is to be understood as the curve (\ref{c+cn}) in the 
parameter space $(E,g)$ (with a definite sign of $g\ne 0$).

\begin{figure}
\begin{center}
\includegraphics[width=14cm,clip=0,angle=0]{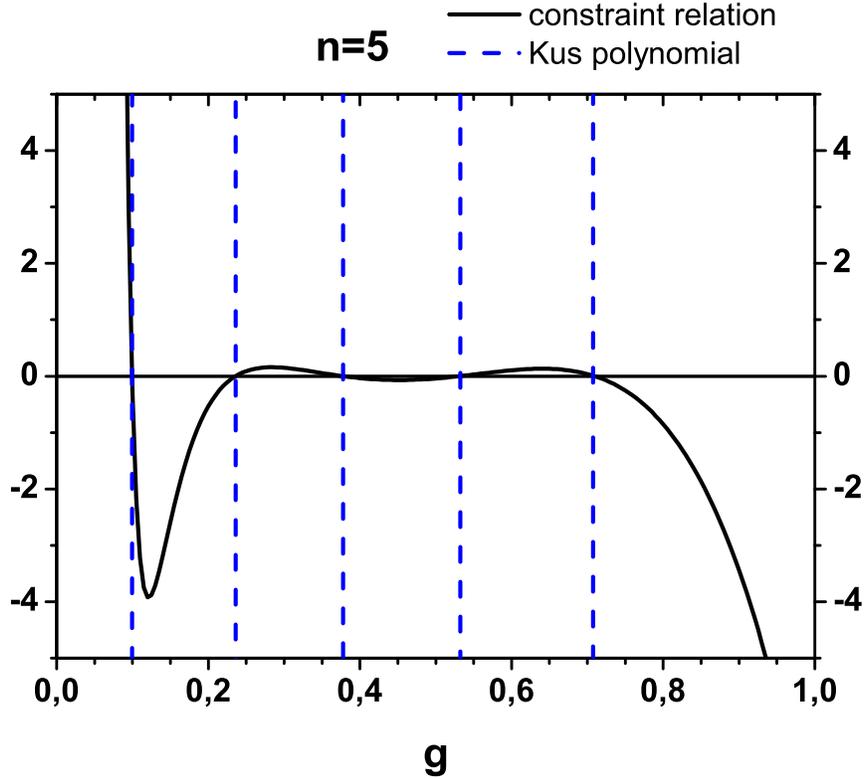}
\end{center}
\caption{A comparison of the polynomial constraint relation (\ref{rmgmrc}) and the Kus polynomial 
for the Rabi model the fifth baseline shows that they have coinciding zeros. 
The polynomials are plotted as a function of $g$ with fixed $\Delta=0.1$ and $\omega=0.4$.
Each zero of the polynomials correspond to the parameter values for which a polynomial solution
of the fifth degree exist. The Kus polynomial has much larger amplitude of oscillations in this case, e.g.
decreasing below $-5\times 10^5$ between the last two zeros. 
Therefore it looks as if crossing $x$ axis vertically.}
\label{fgkus5}
\end{figure}

Focusing now on the $n$th baseline, $b_1(E_n) = (1-2n)\omega^2$,
$c_0(E_n)=n^2\omega^2-\Delta^2-2ng^2$, and
\begin{eqnarray}
&F_1(k) = 2 \omega g (n-k),&
\nonumber\\
&F_0(k) =  k(k-2n) \omega^2 + n^2\omega^2-\Delta^2-2ng^2,
\nonumber\\
&F_{-1}(k) = \frac{ k g}{\omega}(2g^2-\omega^2),\qquad  F_{-2}(k) =-k(k-1) g^2.&
\label{fffr}
\end{eqnarray}
As for any $\gamma=1$ problem, there is an additional single constraint (\ref{rmgmrc}) to be satisfied,
which, when combined with the baseline condition (\ref{c+cn}), yields by Theorem 2 
the necessary and sufficient conditions for the existence of a unique polynomial solution 
of Eq. (\ref{Rabi-diff1}). On substituting (\ref{fffr}) into (\ref{rmgmrc}), the latter becomes
\begin{equation}
P_1(n):= -2 g^2 a_{n2} + \frac{g}{\omega}(2g^2-\omega^2) a_{n1} + 
 \left(n^2\omega^2-\Delta^2-2ng^2\right) a_{n0} =0.
\label{rmgmrc1}
\end{equation}
Because all $F_{\mathfrak g}(k)$ are polynomials in model parameters, the constraint relation (\ref{rmgmrc})
is according to Theorem 3 equivalent to a polynomial identity. For instance,
in the special case of $n=1$ we have $a_{n2}=0$, $a_{n1}=1$. From the system (\ref{gm0rc}) one gets
$a_{10}=-F_0(1)/F_1(0)= (\Delta^2+2g^2)/2g \omega$. The $1$st baseline 
constraint (\ref{rmgmrc1}) then becomes 
\begin{equation}
P_1(1) := \frac{g}{\omega}(2g^2-\omega^2) 
- \left(\omega^2-\Delta^2-2g^2\right) \frac{F_0(1)}{F_1(0)} =
-\frac{ \omega\Delta^2}{2g} \left(\frac{4g^2}{\omega^2}+ \frac{\Delta^2}{\omega^2}-1\right)=0.
\label{rmgmrc11}
\end{equation}
In agreement with Corollary 3, the term in parenthesis is nothing but the Kus polynomial $K_{11}$ 
for the 1st baseline. Indeed, the Kus polynomial $K_{nn}$ associated with the 
$n$th baseline (which is indicated by the left subscript) is defined by 
its own {\em finite} three-term recurrence for each $n\ge 0$ \cite{Ks},
\begin{align}
&K_{n0}=1,\qquad K_{n1}=4\kappa^2+\mu^2 -1,
\nonumber\\
&K_{nl}=(4l\kappa^2+\mu^2 -l^2)K_{n,l-1}
 -4 l(l-1)(n-l+1)\kappa^2K_{n,l-2}.
\nonumber %% \label{ksttrr}
\end{align}
$K_{nn}$ is of $n$th degree (which is indicated by the right subscript) in 
rescaled variables $4\kappa^2$ and $\mu^2$,
where $\kappa=g/\omega$, $\mu=\Delta/\omega$ \cite{Ks}. 
The equivalence of $P_1(n)$ and $K_{nn}$ for higher baseline can be demonstrated numerically, 
which is shown in Fig. \ref{fgkus5}. The only computational difference is that 
the coefficients $a_{n2}$, $a_{n1}$,  $a_{n0}$
entering the constraint polynomial $P_1(n)$ in (\ref{rmgmrc1})
are determined by a {\em four-term downward} recurrence, whereas $K_{nn}$ is obtained by a 
{\em three-term upward} recurrence.
%%%%%%%%%%
%%%%%%%%%%

Given that all zeros of $K_{nn}$ are {\em simple} \cite{Ks}, 
an obvious consequence of the Kus construction \cite{Ks} is that for a given
fixed $0<\mu<1$ there are:
\begin{itemize}

\item [(i)] in total exactly $n$ different polynomial solutions of the Rabi model on each baseline

\item [(ii)] there are no other polynomial solutions of the Rabi model.
\end{itemize}
By varying $\mu$ one selects another set on $n$ points on the $n$th baseline.
For $k<\mu<k+1$, $0<k\le n$, the number
of real zeros of $K_{nn}$ becomes $n-k$ \cite{Ks}. The zeros of $K_{nn}$ correspond
to the points where the energies of positive and negative parity eigenstates cross each other 
as a function of $g$ \cite{Ks}. There are no other degeneracy points of the Rabi model \cite{Ks}.
In particular there is, according to Kus, no $n=0$ baseline solution for $\mu\not\in \mathbb{N}$. 
Obviously Eq. (\ref{Rabi-diff1}) has a $n=0$ solution, but the latter requires
$\Delta=\mu\equiv 0$, in which case $C(z)\equiv 0$. 

In general at the zeros of $K_{nn}$ 
the corresponding wave function component $\psi_+(z)$ is given by
\begin{equation}
\psi_+(z)=e^{-\frac{g}{\omega}z}\prod_{i=1}^n(z-z_i),
\nonumber
\end{equation}
and the component $\psi_-(z)=e^{-gz/\omega}\phi_-(z)$ with $\phi_-(z)$ is determined by 
the first equation of (\ref{Rabi-diff}) for $\Delta\neq 0$.
The differential operator 
\begin{equation}
{\cal L}^1_R\equiv (\omega z+g)\frac{d}{dz}-\left(\frac{g^2}{\omega}+E\right)
\nonumber
\end{equation}
in the first equation of (\ref{Rabi-diff}) is grade zero operator, and hence is exactly solvable, 
because ${\cal L}^1_R$ preserves 
$\mathcal P_n={\rm span}\{1,z, z^2,\ldots,z^n\}$ for any system parameters.
Given that $\phi_-(z)=-\frac{1}{\Delta}{\cal L}^1_R\,\phi_+(z)$, 
$\phi_-(z)$ automatically belongs to the same invariant subspace as $\phi_+(z)$. 
Here the $n$ simple roots $z_i$ are required to satisfy the 
set of the {\em Bethe Ansatz} algebraic equations (\ref{BAEs-general}),
\begin{equation}
\sum_{l\neq i}^n\frac{2}{z_i-z_l}=\frac{2\omega g z_i^2+(2n-1)\omega^2 z_i
 +g(\omega^2-2g^2)/\omega}{(\omega z_i-g)(\omega z_i+g)},~~~~i=1,2,\ldots, n.
\nonumber %% \label{Rabi-solution-BEs}
\end{equation}
When the constraint (\ref{c1-expression}) on $c_0$ now being $c_0(E_n)=n^2\omega^2-\Delta^2-2ng^2$
is substituted into Theorem 4, the following constraint on $\sum_{i=1}^nz_i$ is obtained,
\begin{eqnarray}
c_0(E_n)=n^2\omega^2-2ng^2 - \Delta^2 &=& -n\left[(n-1)\omega^2 +\omega^2 (1-2n)\right] 
+2\omega g\sum_{i=1}^nz_i
\nonumber\\
&=& n^2\omega^2 +2\omega g\sum_{i=1}^nz_i,
\nonumber
\end{eqnarray}
or
\begin{equation}
\Delta^2+2ng^2+2\omega g\sum_{i=1}^nz_i=0.
\label{Rabi-solution-constraint}
\end{equation}
These relations are exactly the same results obtained by the method of
Juddian isolated exact solution \cite{Jd,Ks}.

\subsection{$sl_2$ algebraization}
%%%%%%%%%%%%%%%%%%%%%%%%%%%%%%%%%%%%%%%%%%%%%%%%%%%%%%%%%%%%%%%
The necessary and sufficient condition (\ref{sl2ac}) for $sl_2$ algebraization with spin $j$
satisfying $2j=n$ reproduces the necessary condition (\ref{c+cn})
for the existence of a polynomial solution of $n$th degree, which is the familiar {\em baseline} condition
for the Rabi model \cite{Schw,Ks}. Algebraically, the necessary condition (\ref{c+cn}) ensures that 
the $z^{n+1}$ term disappears and ${\cal L}$ preserves a finite dimensional subspace 
$\mathcal P_n$. Therefore ${\cal L}$ is quasi-exactly
solvable with invariant subspace $\mathcal P_n$. However only zero modes of ${\cal L}$
are relevant physical solutions.

In virtue of (\ref{c+c}), $T_2\in\mathcal U(sl_2)$ requires 
that $C_+=b_2=-2\omega g$ in Eq. (\ref{sl2qes}). Obviously, a different value of $b_2$ implies a
different $T_2\in\mathcal U(sl_2)$.
On substituting for $a_2, b_1$ from Eq. (\ref{Rabi-diff1}), and on using 
the baseline condition (\ref{c+cn}), the first two conditions in (\ref{sl2acb}) are satisfied by
\begin{align}
C_{00} &= \omega^2,
\nonumber\\
C_0 &= \omega^2-2g^2-2E\omega + (2j-1)\omega^2 =-2j\omega^2.
\nonumber
\end{align}
The constant $C_*$ is then easily determined on substituting the above two expressions, 
together with $c_0(E_n)=n^2\omega^2-\Delta^2-2ng^2$, into the last condition of (\ref{sl2acb}).
Hence the Rabi model Hamiltonian becomes an element of $\mathcal U(sl_2)$ on each baseline - 
all that without the necessity of a gauge transformation
usually required for the time-independent Schr\"odinger equation eigenvalue problems.
The point of crucial importance is that with $E$ being fixed by (\ref{c+cn}), 
all the coefficients of the linear 2nd order ODE (\ref{Rabi-diff1}) are {\em fixed}, 
and thus there are only {\em two} linearly independent
solutions of Eq. (\ref{Rabi-diff1}), with at most one of them being polynomial one. 

The same conclusion follows also by using 
the alternative 2nd ODE's of Schweber \cite{Schw} and Koc et al \cite{KKT}.
We find it necessary to check it, because the non Sturm-Liouville
character of underlying equations prevents any kind of a residual
$sl_2$ gauge transformation as employed for conventional Sturm-Liouville problems in Ref. \cite{GKO}.
The results are summarized in Appendix \ref{sec:alt}.

\section{Driven Rabi model}
\label{sc:drm}
%%%%%%%%%%%%%%%%%%%%%%%%%%%%%%%%%%%%%%%%%%%%%%%%%%%%%%%%%
The Hamiltonian of the driven (also called as asymmetric) Rabi model is 
(for more detail see Ref. \cite{Zh6,Zhong14,Li15,Wa,KRW})
\begin{equation}
\hat{H}_R=\omega \mathds{1} a^\dagger a+\Delta\sigma_z
+g\,\sigma_x\left(a^\dagger+a\right)+\delta\,\sigma_x,
\label{RabiHd}
\end{equation}
differs from $\hat{H}_R$ in Eqs. (\ref{rabih}), (\ref{RabiH}) by the 
addition of the driving term $\delta \sigma_x$.
The latter breaks the $Z_2$ symmetry of the Rabi model. 
The driven Rabi model (\ref{RabiHd}) is relevant to the description of some hybrid mechanical 
systems (see e.g. \cite{Zhong14}). With the substitution 
\begin{equation}
\psi_\pm(z)=e^{-gz/\omega}\phi_\pm(z), 
\nonumber
\end{equation}
and on eliminating $\phi_-(z)$ from the system we obtain the uncoupled 
differential equation for $\phi_+(z)$,
\begin{equation}
\hat{H}_R\phi_+(z)=\Delta^2\phi_+(z),
\label{Spectral-eqn-Rabi}
\end{equation}
where \cite{Zh6,Li15}
\begin{eqnarray}
\hat{H}_R &=& (\omega z-g)(\omega z+g)\frac{d^2}{dz^2}+
\bigg[-2\omega gz^2+(\omega^2-2g^2-2E\omega)z-g\omega 
\nonumber\\
& & +2g\left(\frac{g^2}{\omega}-\delta\right)\bigg]\frac{d}{dz}
 +2g\left(\frac{g^2}{\omega} +E-\delta\right)z+E^2-\left(\delta-\frac{g^2}{\omega}\right)^2.
 \label{Rabi-H}
\end{eqnarray}

In contrast to the non-driven Rabi model, the substitution
\begin{equation}
\tilde\psi_\pm(z)=e^{+gz/\omega}\tilde\phi_\pm(z)
\nonumber
\end{equation}
leads to another set of solutions of the driven Rabi model.
On elimination of $\varphi_+(z)$ one obtains the uncoupled differential equation 
for $\varphi_-(z)$,
\begin{equation}
\hat{\tilde H}_R\tilde\varphi_-(z)=\Delta^2\tilde\varphi_-(z),
\label{Spectral-eqn-Rabi'}
\end{equation}
where \cite{Zh6,Li15}
\begin{eqnarray}
\hat{\tilde H}_R &=& (\omega z-g)(\omega z+g)\frac{d^2}{dz^2}+
\bigg[2\omega gz^2+(\omega^2-2g^2-2E\omega)z+g\omega
\nonumber\\
& & -2g\left(\frac{g^2}{\omega}+\delta\right)\bigg]\frac{d}{dz}
 - 2g\left(\frac{g^2}{\omega} +E+\delta\right)z+E^2-\left(\delta+\frac{g^2}{\omega}\right)^2.
\label{Rabi-H'}
\end{eqnarray}
The above two cases are related by a symmetry 
$\psi_+(z,\delta) = \tilde\psi_-(-z,-\delta)$, $\psi_-(z,\delta) = \tilde\psi_+(-z,-\delta)$
in the wave function components \cite{Li15}.

Again one can easily verify that the grading characteristics of the respective differential operators
$\hat{H}_R$ and $\hat{\tilde H}_R$ in Eqs. (\ref{Rabi-H}) and (\ref{Rabi-H'}), respectively, are 
as announced in Eq. (\ref{grmch}): $\gamma=1$, $w=4$, and
\begin{equation}
{\cal F}_1=\mp 2\omega gz^2d_z \pm 2g\left(\frac{g^2}{\omega} +E\mp \delta \right)z,
\qquad
F_1(n)=\mp 2n \omega g\pm 2g\left(\frac{g^2}{\omega}+E\mp \delta\right).
\nonumber
\end{equation}
The baseline condition $F_1(n)=0$ is the
(i) necessary condition for the existence of a polynomial solution
of degree $n$ and (ii) necessary and sufficient condition for $sl_2$ algebraization with spin $j=n/2$.
In each case it implies that the exact (exceptional) energies of the driven Rabi model
are restricted to the baselines \cite{Zh6,Zhong14,Li15},
\begin{equation}
E_n=n \omega-\frac{g^2}{\omega}\pm \delta,~~~~~n=0,1,2,\ldots.
\label{Rabi-solution-E}
\end{equation}

Imposing the constraint (\ref{Rabi-solution-E}) {\em unambiguously} fixes all 
the coefficients of the corresponding 
2nd order linear differential operators $\hat{H}_R$ and $\hat{\tilde H}_R$.
Therefore each of the respective spectral problems 
(\ref{Spectral-eqn-Rabi}) and (\ref{Spectral-eqn-Rabi'}) can have at most two linearly 
independent solutions. The respective spectral problems (\ref{Spectral-eqn-Rabi}) 
and (\ref{Spectral-eqn-Rabi'}) lead to
the anomalous alternative ({\bf A2}) of the ODE (\ref{ODE}) with $B\ne -A'$.
Because the latter violates the necessary condition (\ref{vshc}) 
for Eq. (\ref{ODE}) with fixed polynomial coefficients $A(z), B(z), C(z)$ to have two linearly 
independent (and hence to possess only) polynomial solutions, any polynomial solution
the spectral problems (\ref{Spectral-eqn-Rabi}) and (\ref{Spectral-eqn-Rabi'}) is necessarily {\em unique}.
One arrives at the same conclusion also from that $F_1(n)$ is a {\em linear} function of $n$ and 
on applying Corollary 2.

\begin{figure}
\begin{center}
\includegraphics[width=14cm,clip=0,angle=0]{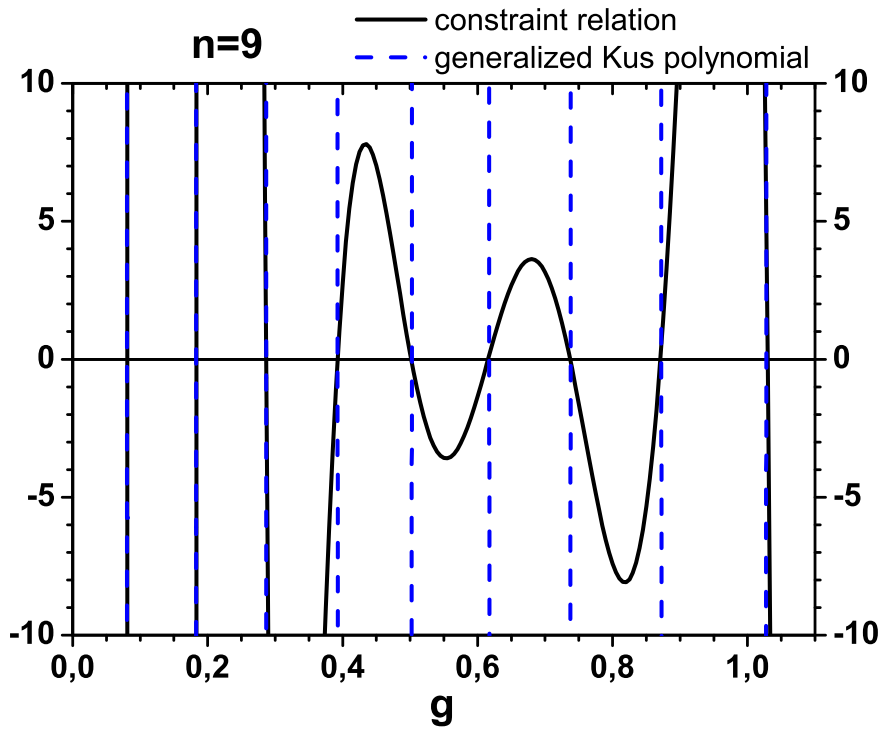}
\end{center}
\caption{A comparison of the polynomial constraint relation (\ref{rmgmrc}) and the generalized 
Kus polynomial (generated by recursion (B.1) of Ref. \cite{Li15}) for 
the driven Rabi model at the ninth baseline shows that 
they have coinciding zeros. The polynomials are plotted 
as a function of $g$ with fixed $\Delta=0.1$, $\omega=0.4$, and $\delta=0.02$.
Each zero of the polynomials corresponds to the parameter values for which polynomial solution
of the ninth degree exist. Around the first three and the last crossings of the $x$-axis the respective 
polynomials appear to be overlapping.}
\label{fggkus9}
\end{figure}

On the $n$th baseline, 
\begin{eqnarray}
b_1 (E_n) &=& \omega^2-2g^2-2E\omega= (1-2n)\omega^2 \mp 2\delta\omega,
\nonumber\\
c_0(E_n) &=&  E^2-\Delta^2- \left(\delta\mp \frac{g^2}{\omega}\right)^2
=n^2\omega^2- \Delta^2 -2n g^2 \pm 2n\omega\delta.
\label{b1c0}
\end{eqnarray}
The driving term $\delta \sigma_x$ in (\ref{RabiHd}) does not change
only the grade $-2$ term of the QRM. 
Therefore we can use only the result for $F_{-2}(k)$ from (\ref{fffr}),
whereas the remaining  coefficients of the recurrence system (\ref{gm0rc}) become
\begin{eqnarray}
&F_1(k) =\pm 2 \omega g (n-k),&
\nonumber\\
&F_0(k) =   k(k-2n) \omega^2 + n^2\omega^2-\Delta^2-2ng^2 \pm 2(n-k)\omega\delta,
\nonumber\\
&F_{-1}(k) =\pm \frac{ k g}{\omega}(2g^2-\omega^2\mp 2\omega\delta),\qquad  F_{-2}(k) =-k(k-1) g^2.&
\label{fffrd}
\end{eqnarray}
Because all $F_{\mathfrak g}(k)$ are polynomials in model parameters, the constraint relation (\ref{rmgmrcd})
is, according to Corollary 3, equivalent to a polynomial identity.
On substituting from (\ref{fffr}) and (\ref{fffrd}) into (\ref{rmgmrc}), the constraint becomes
\begin{equation}
P_1:= -2 g^2 a_{n2} \pm \frac{g}{\omega}(2g^2-\omega^2\mp 2\omega\delta) a_{n1} + 
 \left(n^2\omega^2-\Delta^2-2ng^2\pm 2n\omega\delta\right) a_{n0} =0.
\label{rmgmrcd}
\end{equation}
For instance, in the special case of $n=1$ one has 
$a_{10}=-F_0(1)/F_1(0)=\pm (\Delta^2+2g^2)/2g \omega$. 
The $1$st baseline polynomial constraint becomes
\begin{equation}
\mp \frac{\Delta^2}{2g\omega } 
    \left( 4g^2 + \Delta^2- \omega^2 \pm 2\omega\delta \right)=0,
\label{rmgmrc12}
\end{equation}
which differs from (\ref{rmgmrc11}) in the usual Rabi case by the $\delta$-dependent term.
The term in parenthesis coincides with the generalized Kus polynomials on the $1$st baseline
(cf. recursions (B.1), (B.11) of Ref. \cite{Li15}; recursions (5.1-2) for $\omega=1$ of Ref. \cite{Wa}).
As shown in Fig. \ref{fggkus9} the latter holds for any baseline.
%%%%%%%%%%
%%%%%%%%%%

With $a_2=\omega^2$, the constant $C_*$ required for the $sl_2$ algebraization is easily determined on substituting 
the expressions for $b_1(E_n)$ and $c_0(E_n)$ from (\ref{b1c0}) into the last condition of (\ref{sl2acb}).
Contrary to claims made in Refs. \cite{Zh6,Zhong14,Li15}, we want to 
emphasize that the $sl_2$ algebraization
{\em does not} automatically mean that the corresponding spectral problems (\ref{Spectral-eqn-Rabi}) 
and (\ref{Spectral-eqn-Rabi'})
possess $(n+1)$ eigenfunctions, respectively, in the form of polynomials of degree $n$. 
The coefficient $C_+$ of $T_2\in\mathcal U(sl_2)$ in Eq. (\ref{sl2qes}) 
has to satisfy (\ref{c+c}). Because we have $b_2=\pm 2\omega g$ in Eqs. (\ref{Rabi-H}), (\ref{Rabi-H'}),
a different value of $b_2$ automatically implies a different $T_2\in\mathcal U(sl_2)$.

\section{Nonlinear Rabi model generalizations}
\label{sec:nlrm}
%%%%%%%%%%%%%%%%%%%%%%%%%%%%%%%%%%%%%%%%%%%%%%
Let us consider two different nonlinear Rabi model generalizations.
They are nonlinear in a quantum optics sense, because they describe 
a multi-photon or multi-mode interaction. A corresponding
differential equation remains {\em linear}, which will allows us to 
apply the same strategy as before.

The Hamiltonian of the {\em two-photon} Rabi model reads \cite{AMdn,Zh2,Zh6,NLL,EB1}
\begin{equation}
\hat{H}_{2p}=\omega \mathds{1} a^\dagger a+\Delta\sigma_z+g\,\sigma_x\left[(a^\dagger)^2+a^2\right].
\nonumber %%\label{2-photon-RabiH1}
\end{equation}
The Fock-Bargmann Hilbert space $\mathfrak{B}$ based on the coherent states
associated with the Heisenberg algebra \cite{Brg} gets replaced by a 
more general Hilbert space of entire analytic functions of growth $(1,1)$ associated with 
the so-called Barut-Girardello coherent states \cite{BG} 
of the annihilation operator $K_-$ of the $su(1,1)$ Lie algebra.
With the substitution
\begin{equation}
\psi_\pm(z)=e^{-\frac{\omega}{4g} (1-\Omega) z}\varphi_\pm(z),~~~~~~
\Omega=\sqrt{1-\frac{4g^2}{\omega^2}},
\nonumber %%\label{2-photon-sustitution}
\end{equation}
where $\left|2g/\omega\right|<1$ \cite{NLL}, and on eliminating $\varphi_-(z)$ from the system, 
one obtains the 4th-order differential equation for $\varphi_+(z)$,
\begin{equation}
\hat{H}_{2p}\varphi_+(z)=-\Delta^2\varphi_+(z),
\label{Spectral-eqn-2photon}
\end{equation}
where \cite{Zh6}
\begin{eqnarray}
\hat{H}_{2p} &=& 16g^2 z^2\frac{d^4}{dz^4}+64g^2\left[\frac{\omega}{4g}(\Omega-1) z^2
+\left(q+\frac{1}{2}\right)z\right]\frac{d^3}{dz^3}
\nonumber\\
& &+\left\{4\omega^2(\Omega^2-3\Omega+1)z^2+16\omega g\left[3\left(q+\frac{1}{2}\right)
\Omega-3q-1\right]z+64g^2q\left(q+\frac{1}{2}\right)\right\}\frac{d^2}{dz^2}
\nonumber\\
& &+\left\{\frac{2\omega^3}{g}\Omega(1-\Omega)z^2+\left[8\omega^2q(1-\Omega)+8\omega^2
\left(q+\frac{1}{2}\right)(1-\Omega)^2\right.\right.
\nonumber\\
& &~~~~~\left.\left. +\, 4\omega\left(E-2\omega\left(q+\frac{1}{4}\right)\right)\right]z 
+32\omega gq\left[\left(q+\frac{1}{2}\right)\Omega-q\right]\right\}\frac{d}{dz}
\nonumber\\
& &+\, \frac{\omega^2}{g}(1-\Omega)\left(2q\omega\Omega-\frac{1}{2}\omega-E\right)z
 +4\omega^2q^2(1-\Omega)^2-\left[E-2\omega\left(q-\frac{1}{4}\right)\right]^2.
 \label{2-photon-H1}
\end{eqnarray}
In arriving at the equation, a single-mode bosonic realization of $su(1,1)$ 
has been used. The latter provides 
the infinite-dimensional unitary irreducible representation of
$su(1,1)$ known as the positive discrete series ${\cal D}^+(q)$,
with the quadratic Casimir operator taking the value ${\cal C}=3/16$ and
the so-called Bargmann index $q=1/4, 3/4$ \cite{Zh2,Zh6}.

The Hamiltonian of the nonlinear {\em two-mode} quantum Rabi model reads \cite{Zh2,AMdn,Zh6}
\begin{equation}
\hat{H}_{2m}=\omega \mathds{1} (a_1^\dagger a_1+a_2^\dagger a_2)
+\Delta\sigma_z+g\,\sigma_x(a_1^\dagger a_2^\dagger+a_1 a_2),
\nonumber %%\label{2-mode-RabiH1}
\end{equation}
where one assumes the boson modes to be degenerate with the same frequency $\omega$.
Note in passing that the model is different from 
a {\em linear} two-mode quantum Rabi model studied in Ref. \cite{ChR}.
With the substitution
\begin{equation}
\psi_\pm(z)=e^{-\frac{\omega}{g} (1-\Lambda) z}\varphi_\pm(z),~~~~~~\Lambda=\sqrt{1-\frac{g^2}{\omega^2}},
\nonumber
\end{equation}
where $\left|g/\omega\right|<1$ \cite{NLL}, and on 
eliminating $\varphi_-(z)$ from the system, one obtains the 4th-order differential equation for $\varphi_+(z)$,
\begin{equation}
\hat{H}_{2m}\varphi_+(z)=-\Delta^2\varphi_+(z),
\label{Spectral-eqn-2mode}
\end{equation}
where \cite{Zh6}
\begin{eqnarray}
\hat{H}_{2m}&=&g^2z^2\frac{d^4}{dz^4}+4g^2\left[\frac{\omega}{g}(\Lambda-1) z^2
 +\left(q+\frac{1}{2}\right)z\right]\frac{d^3}{dz^3}
 \nonumber\\
& &+\left\{4\omega^2(\Lambda^2-3\Lambda+1)z^2+4\omega g\left[3\left(q+\frac{1}{2}\right)
\Lambda-3q-1\right]z
 +4g^2q\left(q+\frac{1}{2}\right)\right\}\frac{d^2}{dz^2}
 \nonumber\\
&&+\left\{\frac{8\omega^3}{g}\Lambda(1-\Lambda)z^2+\left[8\omega^2q(1-\Lambda)
 +8\omega^2\left(q+\frac{1}{2}\right)(1-\Lambda)^2\right.\right.
 \nonumber\\
&&~~~~~~\left.\left. +\, 4\omega(E-2\omega q)\right]z
 +8\omega gq\left[\left(q+\frac{1}{2}\right)\Lambda-q\right]\right\}\frac{d}{dz}
 \nonumber\\
&& +\,\frac{4\omega^2}{g}(1-\Lambda)\left(2q\omega\Lambda-\omega-E\right)z
+ 4\omega^2q^2(1-\Lambda)^2-\left[E-2\omega\left(q-\frac{1}{2}\right)\right]^2.
\label{2-mode-H1}
\end{eqnarray}
In arriving at the equation, a two-mode bosonic realization of $su(1,1)$ 
has been employed that requires the quadratic Casimir
to take the value ${\cal C}=q(1-q)$,
where $q>0$ stands for the Bargmann index, which can be any positive integers or half-integers,
i.e. $q=1/2, 1, 3/2,\ldots$ \cite{Zh2,Zh6}.

The point of crucial importance is that those $4$th order linear differential 
operators $\hat{H}_{2p}$ and $\hat{H}_{2m}$ have rather special form.
From the grading point of view, 
the $4$th derivative term has grade $-2$, whereas
the $3$rd derivative contributes two terms with grade $-2$ and $-1$, respectively.
The same as for the (driven) Rabi model, the highest grade terms of each of
$\hat{H}_{2p}$ and $\hat{H}_{2m}$ are the terms $\sim z^2d_z$ and $\sim z$, 
both having a positive grade of $+1$.
Consequently, on using the same notation as in (\ref{ODE}), (\ref{yaoc}) for 
the respective $2$nd order linear differential 
parts of the operators $\hat{H}_{2p}$ and $\hat{H}_{2m}$ 
we have the usual anomalous ODE characterized by the grading parameters as summarized by (\ref{grmch}).
The necessary condition for the existence of a polynomial solution 
of each of Eqs. (\ref{Spectral-eqn-2photon}) and (\ref{Spectral-eqn-2mode}) remains
identical to (\ref{c+cgn}).

In the case of Eq. (\ref{Spectral-eqn-2photon}) the necessary condition (\ref{c+cgn}) becomes 
\begin{eqnarray}
F_1(n):= \frac{\omega^2}{g}(1-\Omega)\left(2q\omega\Omega-\frac{1}{2}\omega-E+ 2n\omega\Omega\right)=0,
\nonumber
\end{eqnarray}
which yields the following baseline constraint on energy,
\begin{eqnarray}
&&E_n=-\frac{1}{2}\, \omega+2(n+q)\omega\Omega,
~~~~~n=0,1,2,\ldots.
\label{2-photon-energy}
\end{eqnarray}
Here it was assumed that $\Omega\neq 1$; the $\Omega=1$ case is trivial corresponding to $g=0$.
Eq. (\ref{2-photon-H1}) yields
\begin{eqnarray}
a_2 &=& 4\omega^2(\Omega^2-3\Omega+1),
\nonumber\\
b_1(E_n) &=& 
 8\omega^2\left(q+\frac{1}{2}\right)(\Omega^2-3\Omega+1) 
 + 4\omega\left(E+\omega\Omega-\frac{1}{2}\,\omega\right)
 \nonumber\\
 &=& 
 8\omega^2\left(q+\frac{1}{2}\right)(\Omega^2-3\Omega+1) 
 + 4\omega\left[-\omega +(2n+2q+1)\omega\Omega\right]
\nonumber\\
 &=& 
 8\omega^2\left[\left(q+\frac{1}{2}\right)\Omega^2 +(n-2q-1)\Omega+q\right],
\nonumber\\
c_0(E_n) &=& 4\omega^2q^2(1-\Omega)^2-\left[E-2\omega\left(q-\frac{1}{4}\right)\right]^2
\nonumber\\
 &=& - 4\omega^2\left[ n^2\Omega^2+2nq \Omega(\Omega-1)\right],
\label{c0ef1}
\end{eqnarray}
where in arriving at the last two expressions we have substituted the baseline condition 
(\ref{2-photon-energy}) for $E_n$. On the $n$th baseline the coefficients of the recurrence 
system (\ref{gm0rc}) are
\begin{eqnarray}
F_1(k) &=& -2(n-k)(1-\Omega)\, \frac{\omega^3 \Omega}{g},
\nonumber\\
F_0(k) &=& k(k-1) a_2 +kb_1(E_n)+c_0(E_n),
\nonumber\\
F_{-1}(k) &=& 16 k(k-1)(k-2) \omega g (\Omega-1)+
16 k(k-1) \omega g\left[3\left(q+\frac{1}{2}\right)
\Omega-3q-1\right]
\nonumber\\
&& +\, 32k\omega gq\left[\left(q+\frac{1}{2}\right)\Omega-q\right],
\nonumber\\
F_{-2}(k) &=& 16 k(k-1)(k-2)(k-3)g^2 
+ 64 k(k-1)g^2 ( k-2+q)  \left(q+\frac{1}{2}\right).
\label{fffr2p}
\end{eqnarray}
The constraint polynomial of Eq. (\ref{rmgmrc}) is determined by
\begin{eqnarray}
& F_0(0) = c_0(E_n), \qquad F_{-1}(1) = 32\omega gq\left[\left(q+\frac{1}{2}\right)\Omega-q\right],&
\nonumber\\
& F_{-2}(2) = 128 g^2 q \left(q+\frac{1}{2}\right).&
\nonumber
\end{eqnarray}

The necessary condition (\ref{sl2ac}) for $sl_2$ algebraization reproduces the 
necessary condition (\ref{c+cgn}) for the existence of a polynomial solution of $n$th degree 
if and only if the spin $j$ of an irreducible $sl_2$ representation satisfies $2j=n$.
Thereby the energies of the 2-photon Rabi model are constraint by the baseline condition (\ref{2-photon-energy}).
The $sl_2$ algebraization conditions are satisfied by
\begin{eqnarray}
C_{00} &=& a_2=4\omega^2(\Omega^2-3\Omega+1), 
\nonumber\\
C_0 &=& b_1 +(2j-1)a_2 = 8\omega^2\left(q+j\right)(\Omega^2-3\Omega+1) + 4\omega^2[(2n+2q+1)\Omega -1].
\nonumber
\end{eqnarray}
The constant $C_*$ is then easily determined on substituting the above two expressions, 
together with $c_0$ from Eq. (\ref{c0ef1}), into the last condition of (\ref{sl2acb}).

Similarly, in the case of Eq. (\ref{Spectral-eqn-2mode}) the necessary condition (\ref{c+cgn}) 
imposes
\begin{eqnarray}
F_1(n):=  \frac{4\omega^2}{g}(1-\Lambda)\left(2q\omega\Lambda-\omega-E+2\omega\Lambda\right)=0,
\nonumber
\end{eqnarray}
which yields the following baseline constraint on energy,
\begin{eqnarray}
&&E_n=-\omega+2(n+q)\omega\Lambda.
\label{2-mode-energy}
\end{eqnarray}
Here $\Lambda\neq 1$, as the $\Lambda=1$ case is trivial corresponding to $g=0$. 

Eq. (\ref{2-mode-H1}) yields
\begin{eqnarray}
a_2 &=& 4\omega^2(\Lambda^2-3\Lambda+1),
\nonumber\\
b_1(E_n) &=& 8\omega^2\left(q+\frac{1}{2}\right)(\Lambda^2-3\Lambda+1) + 4\omega(E+\omega\Lambda),
\nonumber\\
 &=& 8\omega^2\left(q+\frac{1}{2}\right)(\Lambda^2-3\Lambda+1) + 4\omega
 \left[-\omega +(2n+2q+1)\omega\Lambda\right]
 \nonumber\\
 &=& 8\omega^2\left[\left(q+\frac{1}{2}\right)\Lambda^2+(n-2q-1)\Lambda+q\right],
 \nonumber\\
c_0(E_n) &=& 4\omega^2q^2(1-\Lambda)^2-\left[E-2\omega\left(q-\frac{1}{2}\right)\right]^2
\nonumber\\
 &=& - 4\omega^2\left[ n^2\Lambda^2+2nq \Lambda(\Lambda-1)\right],
 \label{c0ef2}
\end{eqnarray}
where in arriving at the last two expressions we have substituted the baseline condition (\ref{2-mode-energy}) for $E_n$. 
On the $n$th baseline the coefficients of the recurrence system (\ref{gm0rc}) are
\begin{eqnarray}
F_1(k) &=& -8(n-k) (1-\Lambda)\,  \frac{\omega^3 \Lambda}{g},
\nonumber\\
F_0(k) &=& k(k-1) a_2 +kb_1(E_n)+c_0(E_n),
\nonumber\\
F_{-1}(k) &=& 4 k(k-1)(k-2) \omega g (\Lambda-1)+
4 k(k-1)\omega g\left[3\left(q+\frac{1}{2}\right)\Lambda-3q-1\right]
\nonumber\\
&& +\, 8 k \omega gq\left[\left(q+\frac{1}{2}\right)\Lambda-q\right],
\nonumber\\
F_{-2}(k) &=& k(k-1)(k-2)(k-3)g^2 
+ 4 k(k-1)g^2 (k-2+q)  \left(q+\frac{1}{2}\right).
\label{fffr2m}
\end{eqnarray}
The constraint polynomial of Eq. (\ref{rmgmrc}) is determined by
\begin{eqnarray}
& F_0(0) = c_0(E_n),\qquad  F_{-1}(1) = 8 \omega gq\left[\left(q+\frac{1}{2}\right)\Lambda-q\right],&
\nonumber\\
&  F_{-2}(2) = 8 g^2 q \left(q+\frac{1}{2}\right).& 
\nonumber
\end{eqnarray}

As before, the necessary condition (\ref{sl2ac}) for $sl_2$ algebraization reproduces the 
necessary condition (\ref{c+cgn}) provided that $2j=n$.
Thereby the energies of the two-mode Rabi model are constraint by the baseline condition (\ref{2-mode-energy}).
The $sl_2$ algebraization conditions are satisfied by
\begin{eqnarray}
C_{00} &=& a_2=4\omega^2(\Lambda^2-3\Lambda+1), 
\nonumber\\
C_0 &=&b_1 +(2j-1)a_2 = 8\omega^2\left(q+j\right)(\Lambda^2-3\Lambda+1) + 4\omega^2[(2n+2q+1)\Lambda -1].
\nonumber
\end{eqnarray}
The constant $C_*$ is then easily determined on substituting the above two expressions, 
together with $c_0$ from Eq. (\ref{c0ef2}), into the last condition of (\ref{sl2acb}).
In virtue of (\ref{c+c}), $T_2\in\mathcal U(sl_2)$ requires that $C_+=b_2$ in Eq. (\ref{sl2qes}).
A different value of $b_2$ in Eqs. (\ref{2-photon-H1}), (\ref{2-mode-H1}) thus automatically implies 
a different $\hat{H}_{2p}$ or $\hat{H}_{2m}\in\mathcal U(sl_2)$.

\section{The generalized Rabi model}
\label{sec:grm}
%%%%%%%%%%%%%%%%%%%%%%%%%%%%%%%%%%%%%%
For the sake of comparison with the results by Tomka et al \cite{TAP} their notation 
is largely adopted here.
The generalized, or also known as an asymmetric, Rabi model described by the Hamiltonian
\begin{equation}
 \hat{H}_{\mathrm{gR}}
 =
 \omega \mathds{1} a^\dag a 
 +
 \Delta \sigma_z
 +
 g_1 \left( a^\dag \sigma_- + a\sigma_+ \right)
 +
 g_2 \left( a^\dag \sigma_+ + a\sigma_- \right)
\nonumber %% \label{gR}
\end{equation}
interpolates between the Jaynes-Cummings (JC) model~\cite{JC} (for $g_2=0$) and the
original Rabi model $g_1=g_2$. Numerous different motivations to consider 
this model have been summarized in Refs. \cite{TAP,AMdn}. The substitution 
\begin{equation}
 \psi_1(z) = e^{-\kappa z}\, S(z), \qquad \kappa \equiv \frac{\sqrt{g_1g_2}}{\omega},
\nonumber %% \label{nudf}
\end{equation}
leads to the following $2$nd order ordinary differential equation \cite{TAP}
\begin{equation}
 \left[ 
 \frac{d^2}{dz^2}
 + 
 \left(
 \sum_{s=1}^{3}\frac{\nu_s}{z-\rho_s} -2\kappa \right) \frac{d}{dz}
 + 
 \frac{C_2(z)}{\prod_{s=1}^{3}(z-\rho_s)} 
 \right] S(z) = 0,
 \label{eq:gRM}
\end{equation}
where $C_2(z)=\sum_{l=0}^2c_l z^l$ is a polynomial of degree $2$.
Obviously, the differential equation (\ref{eq:gRM}) has grade $\gamma=2$.
The constants in Eq.~(\ref{eq:gRM}) are
\begin{align}
 \nu_1 & = -\epsilon + 1, \quad \nu_2 = -\epsilon,
 \quad  \nu_3 = -1,
\nonumber
\end{align}
where
\begin{align}
 \mu &\equiv \frac{\Delta}{\omega}, \qquad
 \lambda_\pm \equiv \frac{g_1^2 \pm g_2^2}{2\omega^2}, \qquad 
e \equiv \frac{E}{\omega}, \qquad
 \epsilon \equiv e + \lambda_+
\label{tapp}
\end{align}
are dimensionless quantities,
\begin{equation}
 \rho_1 = \kappa,
 \qquad
 \rho_2 = -\kappa,
 \qquad
 \rho_3 = \nu,
\nonumber
\end{equation}
are the zeros of $A_3(z)$, and
\begin{equation}
 \nu
 =
 \frac{2 \Delta \sqrt{g_1 g_2}}{g_1^2 - g_2^2}= \frac{\mu}{\lambda_-}\, \kappa.
\label{kpdf}
\end{equation}
The coefficients of $C_2(z)$ are given by \cite{TAP}
\begin{align}
  c_0
  &=
  \kappa
  \left(
     \delta^2 - \epsilon^2 + 2 \epsilon \lambda_+ - \lambda_+^2 + \lambda_+
   + \kappa^2   + \kappa^4 \right)
  \nonumber \\
  &+ \kappa \left( \epsilon-\lambda_+-\kappa^2\right),
  \nonumber  \\
  c_1
  &=
  e(e+1)-\delta^2 + \delta \frac{\lambda_+}{\lambda_-}
  +
  \nu \kappa - \kappa^2 - 2 \nu \kappa \epsilon -  \kappa^4,
 \nonumber \\
  c_2 &= 2 \kappa \epsilon.
  \nonumber
\end{align}

Given that $\sum_j\rho_j=\nu$, $\sum_{j<l} \rho_j \rho_l=-\kappa^2$, and
$\rho_1 \rho_2\rho_3=-\nu\kappa^2$, the differential equation (\ref{eq:gRM}) can be recast as
\begin{eqnarray}
& A_3(z) S'' + B_3(z) S' + C_2(z) S = 0,&
 \label{eq:odegRM}\\
& A_3(z) = (z^2 - \kappa^2)(z-\nu)= z^3-\nu z^2 -\kappa^2 z+\nu\kappa^2, &
\nonumber\\
& B_3(z) = -2\kappa A_3(z) -2\epsilon z^2 +(2\epsilon\nu+\kappa-\nu) z - \kappa (\nu-\kappa) , &
\nonumber
\end{eqnarray}
where, in general, $\deg A_3=\deg B_3=3$, $\deg C_2=2$, i.e. 
the usual anomalous alternative ({\bf A2}). The necessary condition (\ref{gnc}) 
for the existence of a polynomial solution $S_n(z)$ of degree $n$ 
reduces to $F_2(n)=nb_3+c_2=0$, which for
$b_3 =-2\kappa$ and $c_2= 2 \kappa \epsilon$ yields the baseline condition
\begin{equation}
\epsilon =n.
\label{gRnc}
\end{equation}
On the $n$th baseline
\begin{eqnarray}
& {\cal F}_2 = -2\kappa z^3 d_z+ 2n \kappa z^2, \qquad {\cal F}_1 
            = z^3 d_z^2 + 2(\kappa  \nu -n) z^2 d_z  + c_1(n) z,&
\nonumber\\
&  {\cal F}_0 = - \nu z^2 d_z^2 + (2\kappa^3 + 2n\nu+\kappa-\nu)  z d_z  + c_0(n),& 
\nonumber\\
& {\cal F}_{-1}  = - \kappa^2 z d_z^2 - [\kappa (\nu-\kappa)+ 2\nu \kappa^3 ] d_z, 
         \qquad  {\cal F}_{-2} = \nu \kappa^2 d_z^2,& 
\nonumber
\end{eqnarray}
where the argument of $c_0(n)$ and $c_1(n)$ indicates that the baseline 
condition (\ref{gRnc}) has been imposed.
The coefficients of the recurrence system (\ref{gm0rc}) are
\begin{eqnarray}
& F_2(l) = 2(n-l)\kappa, \qquad F_1(l) = l (l-1-2n+2\kappa\nu)+ c_1(n),& 
\nonumber\\
& F_0(l) = l [2\kappa^3 + (2n-l)\nu+\kappa]  + c_0(n),& 
\nonumber\\
&  F_{-1}(l) = - l [\kappa (\nu-\kappa)+(l-1) \kappa^2+ 2\nu \kappa^3 ], \qquad
F_{-2}(l) = l(l-1) \nu \kappa^2.& 
\label{fffrgR}
\end{eqnarray}
The first constraint polynomial $P_1$ in (\ref{gmrc}),
which takes on the form (\ref{rmgmrc}), is determined by
\begin{eqnarray}
& F_0(0) = c_0(n),\qquad  F_{-1}(1) = -\kappa (\nu-\kappa)-2\nu \kappa^3,\qquad  F_{-2}(2) = 2 \nu \kappa^2.& 
\nonumber
\end{eqnarray}
The second constraint polynomial $P_2$ of (\ref{gmrc}) is determined by
\begin{equation}
P_2(n):= a_{n3} F_{-2}(3)+ a_{n2} F_{-1} (2)+ a_{n1} F_0 (1)+ a_{n0} F_1 (0)=0,
\label{rmgmrc2}
\end{equation}
are
\begin{eqnarray}
&  F_1(0) = c_1(n) ,\qquad F_0(1) = 2\kappa^3 + 2n \nu+\kappa-\nu +c_0(n),&
\nonumber\\
&  F_{-1}(2) = - 2 \kappa^2 - 2 [\kappa (\nu-\kappa)+ 2\nu \kappa^3 ], \qquad  F_{-2}(3) = 6 \nu \kappa^2.
\nonumber
\end{eqnarray}
In order to satisfy the constraints (\ref{gmrc}), one has to determine 
common zeros of the polynomials $P_1,\, P_2$.
The latter task amounts to determining zeros of the {\em resultant} of the two polynomials, $R(P_1,P_2)$, 
also known as the {\em eliminant} \cite{MS}. A resultant is important joint invariant of 
two polynomials, which is expressed as a polynomials function of the coefficients of two polynomials.
Thereby the task of finding polynomial solutions of the generalized Rabi model is reduced 
again to an {\em algebraic} equation.

One determines easily that Eq. (\ref{eq:odegRM}) 
{\em cannot} be viewed as a special case of {\em normally ordered} bilinear combination $T_2$
given by Eq. (\ref{sl2qes}) of $sl_2$ generators (\ref{sl2r}).
In order that $\deg P_4=3$ in Eq. (\ref{p432sl2n}), one has to set $C_{++} \equiv 0$. 
Yet with $C_{++} \equiv 0$ one can never have $\deg P_3=3$ in Eq. (\ref{p432sl2n}).
Thus any polynomial solution of Eqs. (\ref{eq:gRM}), (\ref{eq:odegRM})
does {\em not} belong to the $sl_2$ algebraic sector and is, by 
definition, {\em exceptional} \cite{PTr,GKM3}.

A $sl_2$ algebraic sector reappears only in the special case of $\kappa^2=\nu^2$.
The latter is possible if $\mu=\pm \lambda_-$ [cf. Eq. (\ref{kpdf})]. For instance
on taking $\nu=-\kappa$, the three roots $\rho_{1,2,3}$ degenerate into two (namely to $\pm\kappa$),
the polynomial coefficients in Eq. (\ref{eq:odegRM}) factorize as $A_3(z)=(z+\kappa)^2(z-\kappa)$,
$C_2(z)=(z+\kappa)(2\kappa\epsilon z+d_0/\kappa)$, where \cite{TAP}
\begin{equation}
 \frac{d_0}{\kappa}=\epsilon^2+\lambda_+(\lambda_+-2)-\mu^2-\epsilon(2\lambda_+-1)-2\kappa^2 
 - \kappa^4.
\nonumber %% \label{d0}
\end{equation}
Because also \cite{TAP}
\begin{equation}
B_3(z)=(z+\kappa) \left[-2\kappa z^2 -2\epsilon z +2\kappa(\kappa^2+1)\right],
\label{gRB}
\end{equation}
one can factorize out the monomial $(z+\kappa)$ from Eqs. (\ref{eq:gRM}), (\ref{eq:odegRM}).
Thereby the degree of the polynomial coefficients in Eq. (\ref{eq:odegRM})
is reduced by one, so that $\deg A_3=\deg B_3=2$, $\deg C_2=1$. We arrive back
at the usual anomalous alternative ({\bf A2}) with grade $\gamma=1$, where
the single constraint polynomial $P_1$ on the $n$th baseline in (\ref{rmgmrc}) is obtained 
with the recurrence coefficients [cf. Eq. (\ref{fffrgR})] 
\begin{eqnarray}
F_1(l) &=& 2(n-l)\kappa,\qquad F_0(l) = l(l-1-2n) + (d_0/\kappa),
\nonumber\\
F_{-1}(l) &=& 2 l \kappa (\kappa^2+1), \qquad F_{-2}(l) = - l(l-1) \kappa^2.
\label{fffrgRf}
\end{eqnarray}
The necessary and sufficient condition (\ref{sl2ac}) for $sl_2$ algebraization with spin $j$
reproduces the necessary condition (\ref{gRnc}) for the existence of a polynomial solution of $n$th degree,
$\epsilon=n=2j$. The remaining $sl_2$ algebraization conditions are satisfied by
$C_{00}=a_2=1$, $C_0=-2\epsilon +2j-1$, together with the relation (\ref{sl2acb}) for $C_*$.
The coefficient $C_+$ of $T_2\in\mathcal U(sl_2)$ in Eq. (\ref{sl2qes}) 
has to satisfy (\ref{c+c}). Because we have $b_2=-2\kappa$ in Eq. (\ref{gRB}),
a different value of $\kappa$ automatically implies a different $T_2\in\mathcal U(sl_2)$.

\section{Discussion}
\label{sec:disc}
%%%%%%%%%%%%%%%%%%%%%%%%%%%%
The problems defined by Eqs. (\ref{Rabi-diff1}), (\ref{Rabi-H}), 
(\ref{Rabi-H'}), (\ref{sch3.23}), and (\ref{eq:3}):
%%%%%%%%%%%%%%%
\begin{itemize}

\item are {\em not} a standard Heine-Stieltjes problem, because of $\deg P_4=\deg P_3=2$, and $\deg P_2=1$,
i.e. the degrees of $P_n$ are {\em not} strictly decreasing with $n$ and
$\deg P_3\not<\deg P_4$

\item are {\em not} described by a Fuchsian equation
 
\item are {\em not} a standard eigenvalue problem, because ${\cal L}$ contains terms $\sim Ez\, d_z$, $Ez$, $E^2$

\item do not lead to a Sturm-Liouville problem in an equivalent Schr\"odinger equation form, because
then they are described by a nontrivially energy dependent potential.

\end{itemize}
%%%%%%%%%%%%%%%
In the case of a relative motion of two electrons in an external oscillator potential \cite{Tr94,CH}
one had an ordinary eigenvalue problem ${\cal L}\psi=\lambda\psi$ 
(cf. Eq. (5) of Ref. \cite{Tr94}; Eq. (31) of Ref. \cite{CH}).
When the relevant 2nd order ODE of \cite{Tr94,CH} was recast into a Schr\"odinger equation, 
such an equation also did not reduce to a standard Sturm-Liouville eigenvalue problem. 
A solution of the problem involved a so-called {\em coupling constant metamorphosis} 
between energy parameter and other model parameters to have an algebraization. 
However, in the present case of the Rabi model the situation is {\em different} in
that we do not have an ordinary eigenvalue problem ${\cal L}\psi=\lambda\psi$. 
Rather the problem reduces to the determination of zero modes of ${\cal L}$.
A $sl_2$ algebraization amounts merely to the necessary condition for the existence
of polynomial solutions. 

Original result of Zhang (cf. Eqs. (1.8-10) of Ref. \cite{Zh}) was limited to the $2$nd order
ODE's which precluded its application to nonlinear Rabi model generalizations
described by the $4$th order ODE's of Sec. \ref{sec:nlrm}.
In Theorem 4 of Sec. \ref{sc:thrs} we have succeeded to generalize 
the important result of Zhang to to the case of arbitrary $\gamma\ge 0$, which also covers
the nonlinear Rabi model generalizations of Sec. \ref{sec:nlrm}.
With their simple Ans\"{a}tze, which are close in spirit to that of Kus \cite{Ks},
Emary and Bishop \cite{EB1} were not able to find polynomial solutions of 
two-photon Rabi Hamiltonian $\hat{H}_{2p}$ 
that occur at the level-crossings between energy eigenstates
that have {\em different} Bargmann indices. In the latter case one
eigenstate is composed of odd number states, whilst the other is
composed only of even number states. Consequently, no
superposition of these states could lead to a reduction in the
complexity of either wave function and one is unable to find
polynomial solutions at these level-crossings.
Using our constraint relations it is possible 
to give a definite answer if there are polynomial solutions at those level crossings
and also investigate the case of two-mode Rabi Hamiltonian  $\hat{H}_{2m}$. 
This will be dealt with elsewhere.

In Section \ref{sec:grm} is has been shown that the task of finding 
polynomial solutions of the generalized Rabi model amounts to solving algebraic equation  $R(P_1,P_2)=0$, 
where $R(P_1,P_2)$ is the {\em resultant} of the two constraint polynomials $P_1,\, P_2$.
The conditions determining the locations of the exceptional solutions in parameter
space were given in Ref. \cite{TAP} through the Bethe Ansatz equations, which
turned out to be the same as those of the reduced BCS (Richardson) model having three 
degenerate levels of energies $\epsilon_{1,2,3}$ with degeneracies $\nu_{1,2,3}$ respectively \cite{TAP}. 
Similarly to the Kus recipe \cite{Ks}, the conditions of solvability 
of the Bethe Ansatz equations on a $n$th baseline reduced to looking for zeros of a certain 
$n$th order polynomial (cf. Appendix C of Ref. \cite{TAP}). It would be interesting to explore
relation between our algebraic equation $R(P_1,P_2)=0$ and that of Ref. \cite{TAP}.

Recent decade has witnessed a rapid development and understanding
of QES which goes {\em beyond} the initial paradigm of the 
$sl_2$ algebraization. There are many other finite dimensional polynomial spaces
which are {\em not} irreducible modules for the $sl_2$ action, and in
these cases there might be non-Lie algebraic differential operators 
which leave the space invariant \cite{PTr,GKM3}.
In a {\em direct approach} to quasi-exact solvability \cite{GKM3}
more general polynomial spaces are considered and the set of
differential operators that preserve them are investigated without
any reference to Lie algebras. Models were found possessing 
multiple algebraic sectors, which simultaneously allow for both $sl_2$ 
and exceptional polynomial sectors \cite{GKMd}.
Every {\em exceptional} orthogonal polynomial system was proven to be related 
to a classical system by a Darboux-Crum transformation \cite{AGM,Que1,GKM}.
The existence of polynomial solutions in the absence
of any apparent Lie algebra symmetry has been demonstrated for the generalized 
Rabi model by Tomka et al \cite{TAP}. Hence, although the differential equation~(\ref{eq:gRM}) 
is more general than the one corresponding to the usual Rabi model with $g_1=g_2=g$, 
it still has a polynomial solution $S_n(z)$ of degree $n$ \cite{TAP}.

In a general case, the constraint relation(s) (\ref{gmrc}) {\em need not} to be a polynomial
in model parameters. Therefore, there could, in principle, exist polynomial solutions
which eigenenergy is not governed by an algebraic equation.

\section{Conclusions}
\label{sec:conc}
%%%%%%%%%%%%%%%%%%%%%%%%%%%%%%%%%%%%%%
The idea of gradation slicing of ordinary differential equations with polynomial coefficients
has been demonstrated to be an efficient tool for the analysis of polynomial solutions of such equations.  
The necessary condition for a polynomial solution of $n$th degree to exist 
forces energy to a $n$th baseline. Once the constraint relations (\ref{gmrc}) 
on the $n$th baseline can be solved, a polynomial solution is in principle possible 
even in the absence of any underlying algebraic structure. 
Theorem 3 can be viewed as a recipe for recursively generating constraint polynomials 
by a downward recurrence (\ref{gm0rc}), (\ref{gmrc}) with well defined coefficients
for any $\gamma>0$ problem.
We have succeeded in Theorem 4 to generalize the main result of Zhang 
(cf. Eqs. (1.8-10) of Ref. \cite{Zh}) to the case of arbitrary $\gamma\ge 0$.

The theory was illustrated on the examples of various  Rabi models.
For those models, a baseline is known as a Juddian baseline (e.g. in the case of the Rabi model 
the curve described by the $n$th energy level of a displaced harmonic oscillator 
with varying coupling $g$). The corresponding constraint relations were shown to 
(i) reproduce known constraint polynomials for the usual and driven Rabi models and 
(ii) generate hitherto unknown constraint polynomials for the two-mode, two-photon, and generalized 
Rabi models, implying that the eigenvalues of corresponding polynomial eigenfunctions 
can be determined algebraically. Interestingly, the ODE of the above Rabi models were shown to 
be characterized, at least for some parameter range, by the same unique set of grading parameters.
We have not analyzed here a {\em linear} two-mode quantum Rabi model \cite{ChR}, 
intensity-dependent quantum Rabi models \cite{LLN,RLs}, and a two-mode multi-photon 
intensity-dependent Rabi model of Ref. \cite{Lo}, yet one expects similar conclusions to apply.
Although our main motivation came from anomalous problems (the alternative {\bf A2}), 
the concept can be applied universally.

\section{Acknowledgment}
%%%%%%%%%%%%%%%%%%%%%%%%%%
Discussions with A. Eremenko, B. M. Rodr\'{\i}guez-Lara, R. Milson, and 
Y.-Z. Zhang are gratefully acknowledged.

\appendix

\section{$sl_2$ constraints}
\label{cs:slc}
%%%%%%%%%%%%%%%%%%%%%%%%%%%%%%%%
A projectivized {\em differential} operator realization of the $2j+1$ dimensional 
representation of the $sl_2(\mathbb{R})$ algebra of spin $j$
is provided by the operators \cite{KO,GKO,TU,Trb,Trb1,Tams}
\begin{eqnarray}
& J_+ = z^2 \partial_z - 2j z, \quad
J_0 = z \partial_z - j, \quad J_- =\partial_z,&
\nonumber\\
& {} [J_0,J_\pm]=\pm J_\pm, \quad [J_+,J_-]=-2J_0,&
\label{sl2r}
\end{eqnarray}
where $2j=n\ge 0$ is an integer parameter.
$sl_2(\mathbb{R})$ is a {\em graded} algebra with the grading 
of generators (\ref{sl2r}) \cite{Trb1,Tams}:
\begin{equation}
\deg (J_+) = +1,~~~~ \deg (J_0) = 0,~~~~ \deg (J_-) = -1.
\label{e15}
\end{equation}
Hence
\begin{equation}
\deg ( J_+^{n_+} J_0^{n_0} J_-^{n_-})= n_+ - n_-.
\label{e16}
\end{equation}
On using commutation relations, any product of the $sl_2$ generators can be 
brought into a {\em normally ordered} form, where the generators 
in the product are ordered in decreasing grade from the left to the right, plus
a linear combination of the products with a lesser number of generators.
The corresponding {\em normally ordered} quadratic quasi-exactly-solvable differential
operator can be represented as in Eq. (\ref{sl2qes}).
In the normally ordered bilinear combination 
(\ref{sl2qes}) one can always assume that $C_{+-}\equiv 0$.
In a Lie algebra, any product $XY$ can be recast as $XY=\tfrac{1}{2}\{X,Y\}+\tfrac{1}{2}[X,Y]$,
where the antisymmetric (commutator) part is equivalent 
to a {\em redefinition} of the coefficients $C_a$.
If $C_{+-}\ne 0$, one can, on making use of the commutation relations, 
form a symmetric quadratic term, and then apply the Casimir identity,
$
J_0^2 - \tfrac{1}{2} \{J_+,J_-\} = j(j+1).
$
The antisymmetric part is always equivalent to a {\em redefinition} of the coefficients $C_a$.
Hence the number of free parameters is $par (T_2) = 8$ (or $par (T_2) = 7$ if 
the numerical constant term $C_*$ is ignored). 

The corresponding {\em normally ordered} $T_2$ in Eq. (\ref{sl2qes}) can be recast as
(cf. Theorem 2 of Ref. \cite{GKO})
\begin{equation}
T_2= P_4(z)\, \frac{d^2}{dx^2}+ P_3(z)\, \frac{d}{dx} +P_2(z),
\label{nosl2}
\end{equation}
where the polynomials $P_l(z)$ of the $\le l$th order are given by
\begin{eqnarray}
P_4(z) &=& C_{++} z^4 +C_{+0}z^3 + C_{00} z^2 + C_{0-}z + C_{--},
\nonumber\\
P_3(z) &=& -2(2j-1)C_{++}z^3 + [ -(3j-1)C_{+0}+ C_+]z^2
\nonumber\\
 && - [(2j-1)C_{00} -C_0]z - j C_{0-} + C_-,
\nonumber\\
P_2(z) &=& 2j(2j-1) C_{++}z^2 + 2j [j C_{+0}- C_+]z + C_{00}j^2 - C_0j+C_*.
\label{p432sl2n}
\end{eqnarray}
Note in passing that the respective polynomials $P_l(z)$ (i) do not necessarily
have different order (e.g. if $C_{++}=C_{+0}=0$), 
(ii) nor is their order necessarily equal to $j$ (e.g. if $C_{++} =0$). 
One has in general only that (iii) $\deg P_l\le l$ and (iv) $\deg P_l\le \deg P_m$ for $l<m$.

If one compares the polynomials $P_l(z)$ in Eq. (\ref{p432sl2n}) 
with the polynomials $A(z), B(z), C(z)$ in the ODE (\ref{ODE}), (\ref{yaoc}),
then the $sl_2(\mathbb{R})$ realization (\ref{p432sl2n}) of Eq. (\ref{ODE}) implies 
\begin{eqnarray}
&a_4=C_{++}=-\frac{1}{2}\, (2j-1)b_3,\qquad a_3=C_{+0},\qquad a_2=C_{00},&
\nonumber\\
&a_1=C_{0-},\qquad a_0=C_{--},&
\label{zhn1n}
\end{eqnarray}
\begin{eqnarray}
&b_3=-2(2j-1)C_{++}=-2(2j-1)a_4,\qquad b_2=-(3j-1)C_{+0}+ C_+,&
\nonumber\\
&
b_1=- [(2j-1)C_{00} -C_0],\qquad b_0=- j C_{0-} + C_-.&
\label{zhn2n}
\end{eqnarray}

\section{Equations with all solutions being polynomials}
\label{sc:eqps}
%%%%%%%%%%%%%%%%%%%%%%%%%%%%%%%%%%%%%%%%%%%%%%%%%%%%%%%%
For a second order differential equation in general form,
\begin{equation}
A(z) w''(z)+B(z) w'(z)+C(z) w(z)=0,
\label{gfode}
\end{equation}
the Wronski determinant of two solutions is either nonzero or zero everywhere (cf. Chap. 5.2 of Hille \cite{Hi}).
For two linearly independent (not necessarily {\em polynomial}) solutions of (\ref{gfode})
\begin{equation}
W(u_1,u_2)=u_1(z) u_2'(z) - u_1'(z) u_2(z) \ne 0
\label{m5}
\end{equation}
implies that:
\begin{itemize}

\item[(i)] the respective solutions {\em cannot} have a common zero

\item[(ii)] if $u_l(z_0)=0$ for some $l=1,2$ and $z=z_0$, then $u_l'(z_0)\ne 0$, and vice versa. 

\end{itemize}
%%%%%%%%%%%%%%%
Otherwise the Wronski determinant (\ref{m5}) would be zero.
If one of the solutions is a polynomial, the latter condition (ii) prohibits it from having a multiple zero.

The coefficients of (\ref{gfode}) are proportional to the minors of 
the first row of the determinant,
\begin{equation}
\left|
\begin{array}{ccc}
w'' & w' & w
\\
w_1'' & w_1' & w_1
\\
w_2'' & w_2' & w_2
\end{array}
\right|=0,
\nonumber
\end{equation}
and can be determined as \cite{Hi,EG}
\begin{equation}
A:B:C= \left|
\begin{array}{cc}
 w_1' & w_1
\\
 w_2' & w_2
\end{array}
\right|: - \left|
\begin{array}{cc}
w_1'' & w_1
\\
w_2'' & w_2
\end{array}
\right| : \left|
\begin{array}{cc}
w_1'' & w_1' 
\\
w_2'' & w_2' 
\end{array}
\right|.
\nonumber
\end{equation}
An obvious consequence of the above relation is that
the coefficients of (\ref{gfode}) having two {\em polynomial} 
solutions $w_l$ have to be {\em polynomials}.
The other consequence is that
\begin{equation}
B=-A' \Longrightarrow \deg B=\deg A -1.
\label{vshc}
\end{equation}
The consequence can be shown to hold for a linear ODE of any order.
Its proof requires only an elementary formula for the derivative 
of a corresponding Wronskian.

Given a polynomial $A(z)=\prod_{l=1}^n (z-z_{a_l})$ with {\em simple} zeros,
an equation of the form
\begin{equation}
Ay^{\prime\prime}-A'y'+Cy=0
\label{special}
\end{equation}
has a non-trivial polynomial solution $y$ if and only if 
$$C=(-Ay^{\prime\prime}+A'y')/y,$$
with $y$ given by 
\begin{equation}
y(z)=(z-z_1)\ldots(z-z_m),\quad\quad z_l\in\mathbb{C}\backslash\{z_{a_1},\ldots,z_{a_n}\},
\label{igrek}
\end{equation}
where $(z_1,\ldots,z_m)$ are satisfying
\begin{equation}
2\sum_{l\neq k}\frac{1}{z_k-z_l}-\sum_{l=1}^n\frac{1}{z_k-z_{a_l}}=0.
\label{stiltjes}
\end{equation}

\vspace{.1in}

\noindent
{\bf Lemma} (\cite[Lemma 7]{VS}): {\em If equation $(\ref{special})$ has one
non-trivial polynomial solution, then all its solutions are
polynomials}
\vspace{.1in}
(cf. Heine-Stieltjes problem \cite{Heine1878,Stl,Sz,Schhs,Shp,AMu,AGM}).

\section{Singular points at spatial infinity}
\label{sc:spsi}
%%%%%%%%%%%%%%%%%%%%%%%%%%%%%%%%%%%%%%%%%%%%%%%%%%%%%%%%%%%%%
With $\xi=1/z$ one has
\begin{eqnarray}
\frac{d}{dz} &=& \frac{d}{d\xi}\frac{d\xi}{dz}= - \frac{1}{z^2}\frac{d}{d\xi}=- \xi^2 d_\xi,
\nonumber\\
\frac{d^2}{dz^2} &=& \frac{2}{dz^3}\frac{d}{d\xi} + \frac{1}{z^4}\frac{d^2}{d\xi^2}=
2\xi^3d_\xi+\xi^4 d_\xi^2.
\label{xtmx}
\end{eqnarray}
Thereby Eq. (\ref{Rabi-diff1}) is transformed into
\begin{eqnarray}
&&(\omega -g\xi)(\omega +g\xi)\frac{d^2\phi_+}{d\xi^2}
+\left[\frac{ 2(\omega -g\xi)(\omega +g\xi)-b_1}{\xi}
 -b_0 -b_2 \xi^{-2}\right]\frac{d\phi_+}{d\xi}
\nonumber\\
&&~~~~~~~~~~~~~~~~~~
+\left[c_1 \xi^{-3}+ c_0 \xi^{-2} \right]\phi_+ =0.
\label{Rabi-diff1i}
\end{eqnarray}
Because of the singularities $b_2 \xi^{-2}$ and $c_1 \xi^{-3}$, 
Eq. (\ref{Rabi-diff1}) has an {\em irregular} singular point at $z=\infty$. 
Analogous conclusion holds for Eqs. (\ref{sch3.23}), (\ref{eq:3}),
and for the respective
2nd order linear differential operators $\hat{H}_R$ and $\hat{\tilde H}_R$
defined by Eqs. (\ref{Rabi-H}) and (\ref{Rabi-H'}).

\section{Alternative 2nd order ODE}
\label{sec:alt}
%%%%%%%%%%%%%%%%%%%%%%%%%%%%%%%%%%%%%
The alternative 2nd order ODE (3.23) of Schweber \cite{Schw},
\begin{eqnarray}
&&
z(z-\kappa)\frac{d^2\phi}{dz^2}
+\left[ -\kappa z^2 +\left(\kappa^2 -\frac{2\epsilon}{\omega} +1 \right)z 
+ \left(\frac{\kappa \epsilon}{\omega} - \kappa \right) \right]\frac{d\phi}{dz}
\nonumber\\
&&~~~~~~~~~~~~~~~~~~
+\left\{ \frac{\kappa \epsilon}{\omega}\, z + 
\left[\left(\frac{\epsilon}{\omega}\right)^2-\left(\frac{\Delta}{\omega}\right)^2 
 - \frac{\kappa^2 \epsilon}{\omega} \right]\right\} \phi =0,
\label{sch3.23}
\end{eqnarray}
where $\kappa=2g/\omega$ and energy $\epsilon$ in (\ref{sch3.23}) corresponds to $E+\omega\kappa^2/4$. 
The grading characteristics of the 2nd order ODE are again
the same as announced earlier in Eq. (\ref{grmch}),
and the equation corresponds to the anomalous alternative ({\bf A2}) 
with $\deg A=\deg B=2$, $\deg C=1$, characterized by
\begin{equation}
b_2 = -\kappa,\qquad c_1 =\frac{\kappa \epsilon}{\omega}\cdot
\nonumber
\end{equation}
Hence for Eq. (\ref{sch3.23}) the very same 
$sl_2$ algebraization condition (\ref{sl2ac}) applies, which for $n=2j$ 
coincides with the necessary condition (\ref{c+cgn}) for the existence of a polynomial solution 
of Eq. (\ref{sch3.23}). The condition (\ref{c+cgn}) yields again 
the familiar {\em baseline} condition for the Rabi model,
\begin{equation}
\frac{\kappa \epsilon}{\omega}=n \kappa \Longleftrightarrow \epsilon=n\omega
\Longleftrightarrow
E=n \omega - \frac{g^2}{\omega}\cdot
\nonumber
\end{equation}
We have on the $n$th baseline $a_2=1$ and
\begin{equation}
b_1(\epsilon_n) = \kappa^2 -2n+1,\qquad
c_0(\epsilon_n) = n^2-\mu^2-n \kappa^2.
\nonumber
\end{equation}
Eq. (\ref{c1-expression}) in the case of $\gamma=1$ then yields
\begin{equation}
\mu^2 +\kappa \sum_{i=1}^nz_i=0,
\nonumber
\end{equation}
i.e. an $n$-independent constraint on $\sum_{i=1}^nz_i$ [cf. Eq. (\ref{Rabi-solution-constraint})].

Let us now examine the second order differential equation (5) of Ref. \cite{KKT},
\begin{eqnarray}
z(1-z)\frac{d^2\Re (z)}{dz^2}+\left[ 4 \kappa^2 z^2-(2\kappa^2-2\epsilon+1)z
+ 1-\epsilon-\kappa^2 \right] \frac{d\Re (z)}{dx} 
&& \nonumber \\
+\left[ -4\kappa^2(\epsilon+\kappa^2)z + 3\kappa^4+2\epsilon\kappa^2 -\epsilon^2+\mu^2 \right] 
\Re (z) &=& 0,
\label{eq:3}
\end{eqnarray}
where $\kappa=g/\omega$. Eq. (\ref{eq:3}) corresponds to a different choice of the independent variable
in Eq. (\ref{Rabi-diff1}) and the dimensionless energy $\epsilon$ in (\ref{eq:3}) corresponds 
to $E/\omega$ in (\ref{Rabi-diff1}).

The grading characteristics of the 2nd order ODE are again
the same as announced earlier in Eq. (\ref{grmch}),
and Eq. (\ref{eq:3}) corresponds again to the anomalous alternative ({\bf A2}) 
with $\deg A=\deg B=2$, $\deg C=1$, 
characterized by
\begin{eqnarray}
b_2 &=& 4\kappa^2,\qquad b_1 = 2\epsilon-2\kappa^2-1,
\nonumber\\
c_1 &=& -4\kappa^4-4\epsilon\kappa^2=-4\kappa^2(\epsilon+\kappa^2)
\nonumber\\
c_0 &=& 3\kappa^4+2\epsilon\kappa^2 -\epsilon^2+\mu^2.
\label{kktcf}
\end{eqnarray}
The necessary and sufficient condition (\ref{sl2ac}) for $sl_2$ algebraization 
with spin $j$, or the necessary condition (\ref{c+cgn})
for the existence of a polynomial solution of $n=2j$th degree, yields 
(cf. Eq. (12) of Ref. \cite{KKT})
\begin{equation}
-4\kappa^2(\epsilon+\kappa^2) = -4n\kappa^2 ~~\mbox{or}~~\epsilon_n=n-\kappa^2,\qquad n\ge 0.
\nonumber
\end{equation}
On recalling the rescaling of energy, this
is again the familiar {\em baseline} condition for the Rabi model.

We have on the $n$th baseline
\begin{equation}
c_0(\epsilon_n) = 4n\kappa^2 -n^2+\mu^2,\qquad
b_1(\epsilon_n) = 2n-4\kappa^2-1.
\nonumber
\end{equation}
Eq. (\ref{c1-expression}) in the case of $\gamma=1$ then, given $a_2=-1$, yields
\begin{equation}
4n\kappa^2 -n^2+\mu^2 = - 4\kappa^2 \sum_{i=1}^nz_i+ n(n-1)-n(2n-4\kappa^2-1),
\nonumber
\end{equation}
i.e. again an $n$-independent constraint on $\sum_{i=1}^nz_i$
[cf. Eq. (\ref{Rabi-solution-constraint})],
\begin{equation}
\mu^2 + 4\kappa^2 \sum_{i=1}^nz_i=0.
\nonumber
\end{equation}

%%%%%%%%%%%%%%%%%%%

\end{document}